\documentclass[fleqn,usenatbib]{mnras}

\usepackage{newtxtext,newtxmath}
\usepackage[T1]{fontenc}

\DeclareRobustCommand{\VAN}[3]{#2}
\let\VANthebibliography\thebibliography
\def\thebibliography{\DeclareRobustCommand{\VAN}[3]{##3}\VANthebibliography}

\usepackage{graphicx}	% Including figure files
\usepackage{amsmath}	% Advanced maths commands
\usepackage{tabularx}
\usepackage{multirow}
\usepackage{dcolumn}
\usepackage{subfigure}
\usepackage[export]{adjustbox}
\usepackage{verbatim}
\usepackage{xcolor}
\usepackage{array}
\usepackage{float} %required for the placement specifier H
\usepackage{wasysym}

\DeclareMathOperator\erf{erf}

\newcolumntype{s}{>p{30pt}}
\renewcommand{\arraystretch}{1.3}
\newcommand{\hMpc}{h^{-1}\mathrm{Mpc}}
\newcommand{\wptwo}{w_\mathrm{p}^{(2)}}
\newcommand{\wpthree}{w_\mathrm{p}^{(3)}}
\newcommand{\xitwo}{\xi^{(2)}}
\newcommand{\xithree}{\xi^{(3)}}
\newcommand{\wcen}{w^{c}}
\newcommand{\wsat}{w^{s}}

\title[High order HOD]{Constraining galaxy-halo connection with high-order statistics}

\author[H. Zhang et al.]{
Hanyu Zhang,$^{1}$\thanks{E-mail: hanyuz@phys.ksu.edu (HYZ)}
Lado Samushia,$^{1}$\thanks{E-mail: lado@phys.ksu.edu (LS)}
David Brooks,$^{2}$
Axel de la Macorra,$^{3}$
Peter Doel,$^{2}$
Enrique Gaztañaga,$^{4,5}$
\newauthor
Satya Gontcho A Gontcho,$^{6}$
Klaus Honscheid,$^{7,8}$
Robert Kehoe,$^{9}$
Theodore Kisner,$^{6}$
Aaron Meisner,$^{10}$
\newauthor
Claire Poppett,$^{11}$
Michael Schubnell,$^{12}$
Gregory Tarle,$^{12}$
Kai Zhang,$^{6}$
Hu Zou$^{13}$
\\
$^{1}$Department of Physics, Kansas State University, 116 Cardwell Hall, Manhattan, KS 66506, USA\\
$^{2}$Department of Physics \& Astronomy, University College London, Gower Street, London, WC1E 6BT, UK\\
$^{3}$Instituto de F\'{i}sica, Universidad Nacional Aut\'{o}noma de M\'{e}xico, Cd. de M\'{e}xico C.P. 04510, M\'{e}xico\\
$^{4}$Institute of Space Sciences (ICE, CSIC), 08193 Barcelona, Spain\\
$^{5}$Institut d\'~Estudis Espacials de Catalunya (IEEC), 08034 Barcelona, Spain\\
$^{6}$Lawrence Berkeley National Laboratory, 1 Cyclotron Road, Berkeley, CA 94720, USA\\
$^{7}$Department of Physics, The Ohio State University, 191 West Woodruff Avenue, Columbus, OH 43210, USA\\
$^{8}$Center for Cosmology and AstroParticle Physics, The Ohio State University, 191 West Woodruff Avenue, Columbus, OH 43210, USA\\
$^{9}$Department of Physics, Southern Methodist University, Dallas, TX 75275, USA\\
$^{10}$NSF's National Optical-Infrared Astronomy Research Laboratory, 950 N. Cherry Avenue, Tucson, AZ 85719, USA\\
$^{11}$Space Sciences Laboratory (SSL), UC Berkeley, 7 Gauss Way, Berkeley, CA 94720, USA\\
$^{12}$Physics Department, University of Michigan Ann Arbor, MI 48109, USA\\
$^{13}$Key Laboratory of Optical Astronomy, National Astronomical Observatories, Chinese Academy of Sciences, Beijing
100012, China
}

\date{Accepted XXX. Received YYY; in original form ZZZ}

\pubyear{2022}

\begin{document}
\label{firstpage}
\pagerange{\pageref{firstpage}--\pageref{lastpage}}
\maketitle

\begin{abstract}
We investigate using three-point statistics in constraining the galaxy-halo connection. We show that for some galaxy samples, the constraints on the halo occupation distribution parameters are dominated by the three-point function signal (over its two-point counterpart). We demonstrate this on mock catalogs corresponding to the Luminous Red Galaxies (LRGs), Emission-Line Galaxies (ELG), and quasars (QSOs) targeted by the Dark Energy Spectroscopic Instrument (DESI) Survey. The projected three-point function for triangle sides less up to 20$h^{-1}$ Mpc measured from a cubic Gpc of data can constrain the characteristic minimum mass of the LRGs with a precision of $0.46$ \%. For comparison, similar constraints from the projected two-point function are $1.55$ \%. The improvements for the ELGs and QSOs targets are more modest. In the case of the QSOs it is caused by the high shot-noise of the sample, and in the case of the ELGs, this is caused by the range of halo masses of the host halos. The most time-consuming part of our pipeline is the measurement of the three-point functions. We adopt a tabulation method, proposed in earlier works for the two-point function, to reduce significantly the required compute time for the three-point analysis. 
\end{abstract}


%%%%%%%%%%%%%%%%%%%%%%%%%%%%%%%%%%%%%%%%%%%%%%%%%%

%%%%%%%%%%%%%%%%% BODY OF PAPER %%%%%%%%%%%%%%%%%%

\section{Introduction}

% explain galaxy-halo connection

Simulations of structure formation have become invaluable in analyzing cosmological data \citep{1998ARA&A..36..599B,2020NatRP...2...42V}. They are used for studying nonlinear gravitational evolution, validating and calibrating theoretical models of structure formation, and estimating covariance matrices of clustering measurements. Cold dark matter simulations are the easiest to produce. They provide us with an accurate picture for the clustering of dark matter halos \citep{2005CSci...88.1088B,2011EPJP..126...55D}. The positions of galaxies cannot be obtained from the cold dark matter simulations. They depend on baryonic physics that is not captured by the cold dark matter simulations \citep{2014MNRAS.444.1518V,2015MNRAS.446..521S}. In addition, resolving galaxies in large volumes requires a much higher mass resolution that cannot be realized with current computers.
Galaxy surveys, on the other hand, measure positions of galaxies rather than their host dark matter halos. It is essential to have an accurate method of placing galaxies in these dark matter simulations for the robust analysis of such data.

% explain HOD, mention alternatives

The Halo Occupation Distribution (HOD) approach is currently one of the most widely used methods to achieve this goal \citep{1998ApJ...494....1J,2000MNRAS.318..203S,2000MNRAS.318.1144P,2001ApJ...546...20S,2002ApJ...575..587B,2002PhR...372....1C,2005ApJ...633..791Z,2007ApJ...667..760Z,2009ApJ...707..554Z}. In the HOD framework galaxies are placed in halos based on some probabilistic prescription that depends on the properties of the host halo and its neighborhood. In the basic HOD models, the probability of a halo to host a certain number of galaxies only depends on its mass. In more complicated models it can also depend on the local density of halos around the host and some features of the history of the halo formation. Models of various complexity have been offered for where exactly to place the galaxies inside the halo and how to assign velocities to those galaxies. 

An alternative approach to connect galaxies and halos is the sub-halo abundance matching (SHAM) method \citep{2004ApJ...609...35K,2004MNRAS.353..189V,2006MNRAS.371.1173V,2006ApJ...647..201C,2010ApJ...717..379B,2016MNRAS.459.3040G}. By assuming a monotonic relation between certain halo properties and certain galaxy properties, a galaxy catalog can be generated by matching the observed list of galaxies sorted by galaxy property with a list of halos (and sub-halos) sorted by halo property from simulations. The method based on a Conditional Luminosity Function which models galaxies as a function of both their luminosity and host halo mass is another alternative \citep{2003MNRAS.339.1057Y,2006MNRAS.365..842C,2006MNRAS.369.1869C,2010arXiv1010.5484W}.

% describe how HOD is constrained

The HOD models have adjustable parameters that are tuned to obtain galaxies as similar as possible to the observed sample. Traditionally, they are constrained by their 2-Point Correlation Function (2PCF), which is the likelihood of finding a pair of galaxies with a certain separation. The 2PCF for separations up to 20 $\hMpc$ is usually used for this purpose \citep{2011ApJ...728..126W,2012ApJ...755...30R,2017ApJ...848...76Z,2020MNRAS.497..581A,2020MNRAS.499.5486A,2021MNRAS.505..377R,2021MNRAS.501.3309Z}.  

% refer to previous HOD work

% explain why 3-pt is interesting

The 2PCF alone does not always have enough constraining power. Many different combinations of HOD parameters may result in a 2PCF that is consistent with the data within the measurement errors. One way of improving the constraints is to also fit the observed 3-Point Correlation Function (3PCF), which is a probability of finding a triplet of galaxies with certain side lengths and orientation concerning the line-of-sight with respect to an observer \citet{2018MNRAS.476..814H,2017MNRAS.465.2225H}.

The usage of the 3PCF to constrain galaxy-halo connection has a long history \citep{2004MNRAS.353..287W,2005MNRAS.361..824G,2005ApJ...632...29F,2008ApJ...672..849M}. In more recent works, \citet{2007MNRAS.378.1196K} studied the shape dependence of reduced 3PCF and find that signal from reduced 3PCF could help break the degeneracy between HOD parameters. 
\citet{2011ApJ...737...97M} measured the redshift-space 3PCF of LRGs from SDSS on large scales up to $~ 90 \hMpc$ and use the 3PCF to constraint bias parameters, which in turn help estimate the LRG HOD parameters.
\citet{2015MNRAS.449L..95G} explored the constraining power of redshift space 3PCF on HOD parameters including the galaxy velocity bias.
\citet{Yuan_2018} tested the potential extra constraining power of HOD parameters from squeezed 3PCF \citep{Yuan_2017}.
% describe previous 3-pt work

% explain why 3-pt is hard

\begin{figure*}
\centering
\subfigure {\includegraphics[width=\linewidth]{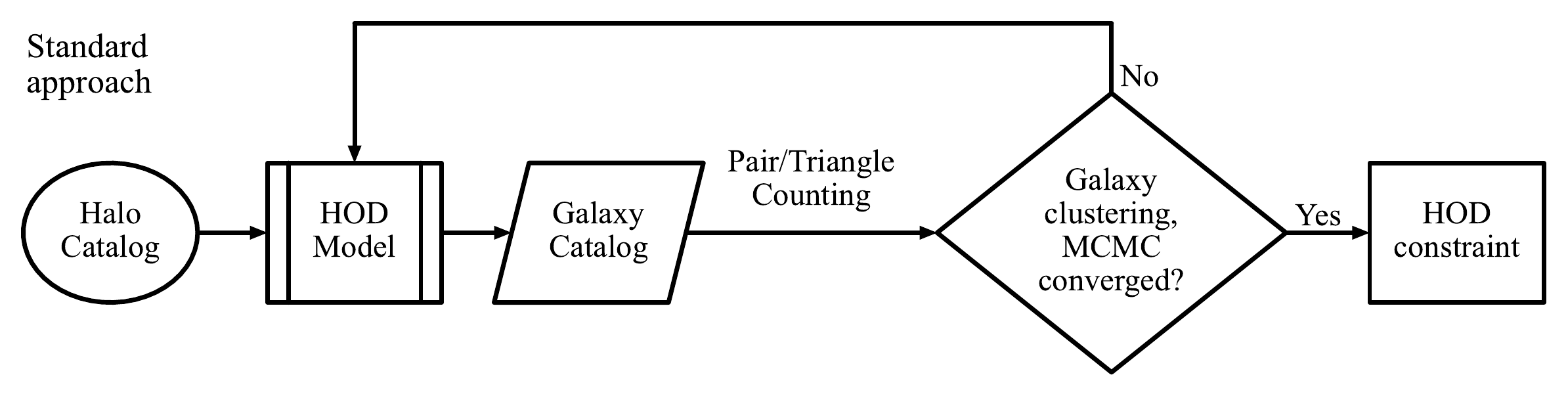}}
\subfigure {\includegraphics[width=\linewidth]{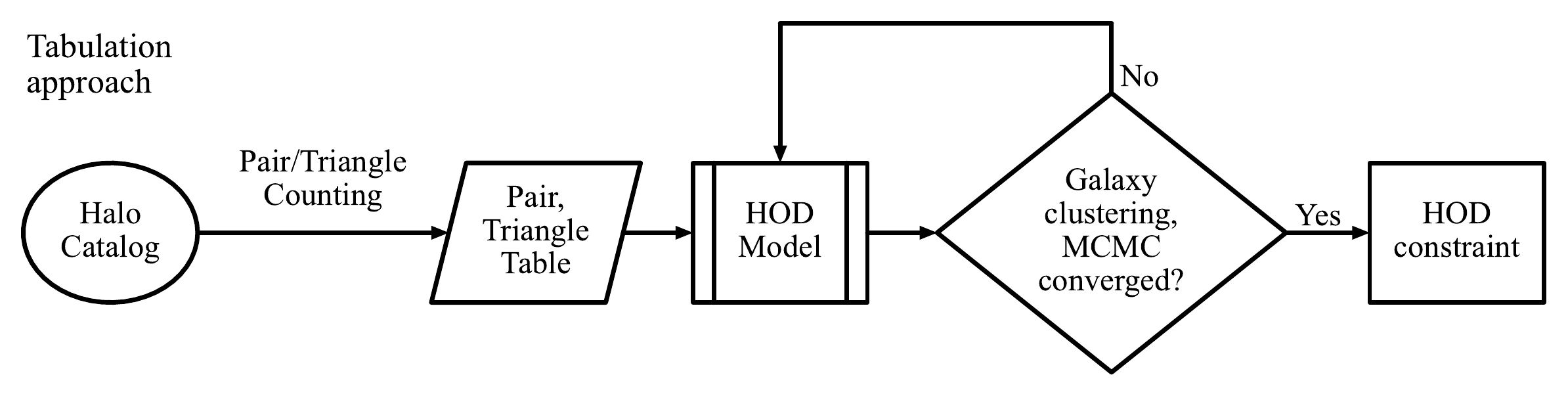}}
\caption{\protect\footnotesize The flow chart on top shows the conventional sequence of steps leading to the HOD constraints. The bottom panel shows the same flow chart for the tabulation approach.}
\label{fig:flowchart}
\end{figure*}

The top diagram on Fig.~\ref{fig:flowchart} schematically shows the steps required to constrain the HOD parameters with a 2PCF or a 3PCF. For a set of HOD parameters we populate mocks with galaxies according to that model, we then measure the clustering statistics of interest, it is compared with a similar measurement from the data, and the posterior likelihood is assessed. This process is repeated many times for various HOD parameter sets until the posterior likelihood is well explored. The most time-consuming part of this algorithm is computing the 2PCF and the 3PCF. Computing the three-point correlation function is especially time-consuming. The number of all possible triplets scales as $N_\mathrm{gal}^3$, where $N_\mathrm{gal}$ is the number of galaxies in the sample. For a big sample, this requires looking at many millions of triangular configurations. This computation needs to be performed at each point in the MCMC chain. Recent works have proposed algorithms that make it possible to compute certain combinations of 3PCF with $N_\mathrm{gal}^2$ complexity \citep{2015MNRAS.454.4142S,2016MNRAS.455L..31S,2022MNRAS.509.2457P}, but even with these algorithmic improvements, this step remains the most computationally expensive piece in the pipeline. 

% explain TabCorr idea
%\begin{figure}
%\centering
%\subfigure {\includegraphics[width=\columnwidth]{flowchart_sta.pdf}}
%\subfigure {\includegraphics[width=\columnwidth]{flowchart_tab.pdf}}
%\caption{\protect\footnotesize The flow chart on top shows the conventional sequence of steps leading to the HOD constraints. The bottom panel shows the same flow chart for the tabulation approach.}
%\label{fig:flowchart}
%\end{figure}

The bottom diagram on Fig.~\ref{fig:flowchart} shows similar schematics for the tabulation approach that has been first proposed in \citet{2011MNRAS.416.1486N} and extended by \citet{2016MNRAS.458.4015Z}. In this approach, the 2PCF of some subsets of halos are precomputed separately before the MCMC stage. These measurements are then combined with certain weights to statistically emulate various HOD population schemes. We describe the tabulation method in detail in Sec.~\ref{sec:tabcorr}. This approach saves a lot of computation time since the most time-consuming part of the algorithm is performed only once before launching the MCMC chain.

% shortcomings of TabCorr
The tabulation method was initially developed for the 2PCF-based fits but it is trivially generalizable to the 3PCF. Many 3PCF-based results that we present in this paper would have required prohibitive computation times with the traditional approach. 
%We generalize the tabulation method to make it applicable to the HOD prescriptions in which the probabilities of galaxies to occupy a certain halo are conditional on other galaxies present in the same halo, including the multi-tracer HODs where the probability of a certain galaxy to be present in a halo is conditional on the presence of other galaxy types (see section \ref{sec:tabcorr} for more details).

% describe basic results and put them in DESI context

We test our method on the galaxies designed to emulate the Luminous Red Galaxies (LRGs), the Emission line galaxies (ELGs), and the Quasars (QSOs) targeted by the Dark Energy Spectroscopic Survey \citep{2016arXiv161100036D}. We show the 3PCF constraints on the HOD parameters dominate the 2PCF results for the DESI-like LRGs. 3PCF has up to 70 $\%$ improvement for a certain parameter. For the ELG and the QSO galaxies, the improvements offered by adding the 3PCF are more modest because of the lower typical host halo mass and lower density of those tracers.

\section{HOD analysis pipeline}
\subsection{HOD model}
\label{sec:baseline} % used for referring to this section from elsewhere

We use a HOD prescription in which the expectation value of galaxies hosted by a dark matter halo only depends on the virial mass of the halo. The expectation value is different for central galaxies that occupy the center of the halo, and satellites that are in virial motion around the center. 

For the LRGs we use
\begin{align}
\langle N_{\mathrm{lrg}}^{c}\rangle(M) &= \frac{A_c}{2}\left(1+\erf\left[\frac{\log(M)-\log(M_{\rm cut})}{\sigma}\right]\right). \label{eq:Ncentlrg}\\
\langle N_{\mathrm{lrg}}^{s}\rangle(M) &= A_s\left(\frac{M - M_0}{M_1}\right)^\alpha H(M-M_0). \label{eq:Nsatlrg}
%\end{split}
\end{align}
The central probability increases with mass until it saturates to some high mass value. The satellite probability is zero below some threshold mass but increases as a power law above that mass.

In both formulas, $M$ is the mass of the host halo. $A_c$, referred to as a duty cycle in the literature, is a maximum probability for high mass halos to host an LRG. $M_\mathrm{cut}$ is the characteristic minimum mass to host an LRG. $\sigma$ describes how steeply the probability increases with halo mass around $M_\mathrm{cut}$. $M_0$ is a mass threshold for the satellite galaxies. $\alpha$ controls the steepness of the increase in the satellite probability with the host halo mass. $M_1$ is the extra mass above the threshold that the halo must have for the expected number of satellites to be equal to one. $A_s$ sets an overall amplitude of the probability. In principle, this parameter is fully degenerate with $M_1$. We use $A_s$ for convenience when creating mock catalogs because it can be changed independently of other parameters to adjust the overall number density of the galaxies without affecting their distribution across masses. $H$ is a Heaviside step function.

This model has been demonstrated to describe well the LRGs in BOSS and eBOSS surveys \citep[e.g.][]{2011ApJ...728..126W,2017ApJ...848...76Z,2020MNRAS.497..581A,2021MNRAS.505..377R,2021MNRAS.501.3309Z}.

For the ELGs \citep[e.g.][]{2020MNRAS.499.5486A} the central probability is a Gaussian function that decays at both high and low mass ends. The satellite probability is similar to the LRGs.
\begin{align}
    \langle N_{\mathrm{elg}}^{c}\rangle(M) &= \frac{A_c}{\sqrt{2\pi}\sigma} \exp{\left(-\frac{\left[\log(M)-\log(M_{\rm cut})\right]^2}{2\sigma^2}\right)} \\
    \langle N_{\mathrm{elg}}^{s}\rangle(M) &= A_s\left(\frac{M-M_{\rm 0}}{M_{\rm 1}}\right)^\alpha H(M - M_0).\label{PsateELG}
\end{align}
$M_\mathrm{cut}$, in this case, describes the most probable halo mass to host a central ELG and $\sigma$ is the variance in the width of this pdf as a function of mass.

For the QSOs, we use a similar formula for the central probability but a slightly modified formula for a satellite probability.
\begin{align}
    \langle N_{\mathrm{qso}}^{c}\rangle(M) &= \frac{A_c}{2}\left(1+\erf\left[\frac{\log(M)-\log(M_{\rm cut})}{\sigma}\right]\right) ,\label{PcentQSO} \\
    \langle N_{\mathrm{qso}}^{s}\rangle(M) &= A_s\left(\frac{M}{M_{\rm 1}}\right)^{\alpha}\exp{\left(-\frac{M_{\rm 0}}{M}\right)}. \label{PsateQSO}
\end{align}
The difference from the LRG is that the QSO hosting probability decays exponentially at lower masses instead of having a sharp cutoff. $M_0$, in this case, controls the decay rate as we go to the lower masses, while $M_1$ set the normalization \citep[e.g.][]{2012ApJ...755...30R,2020MNRAS.499..269S}. 

%We will consider models in which the probability of a satellite is conditioned on whether the specific halo has a central galaxy in it. E.g. we will study cases where the satellite galaxy can only occupy a halo if there is already a central galaxy in that same halo, $N_s \sim N_c \mathrm{Pois}({p^s})$. $N_c$ on the right hand side of the equation is an outcome of the previous random draw for the central (not an independent draw from the same distribution).

There is substantial evidence that the probability of a halo hosting a certain galaxy may depend on other parameters in addition to the virial mass, a phenomenon called an assembly bias \citep{2007MNRAS.374.1303C,2005MNRAS.363L..66G,2017A&A...598A.103P,2018MNRAS.480.3978A,2018ApJ...853...84Z,2020MNRAS.493.5506H,2021MNRAS.501.1603H}. In this work we ignore the assembly bias. This does not affect our main conclusions, since the main objective of our work is to study a potential improvement in the HOD parameter constraints and we don't expect our conclusions to be sensitive to the exact nature of the HOD model. 

\subsection{Mock galaxy catalog}
\label{sec:mock}
We use the \textsc{AbacusSummit} cosmological N-body simulation to create mock galaxy catalogs \citep{2021MNRAS.508..575G, 2021arXiv211011409B,2019MNRAS.485.3370G,2018ApJS..236...43G,2016MNRAS.461.4125G,2009PhDT.......175M}. \textsc{AbacusSummit} were designed to meet the cosmological simulation requirements of DESI. Specifically, we use the \texttt{AbacusSummit\_highbase\_c000\_ph100} box of \textsc{AbacusSummit} with Planck 2018 cosmology, box size of 1000 $\hMpc$ per side, and $3456^3$ dark matter particles with the mass of $2.1\times 10^{9}$ $h^{-1}M_{\odot}$ per particle. We use cleaned \textsc{CompaSO} \citep{2021MNRAS.tmp.2718H} halo catalog at the $z=0.8$, $z=1.1$, and $z=1.4$ snapshots to create the LRG, ELG, and QSO samples respectively. These are the redshifts at which the number densities of the tracers are expected to peak.\textsc{AbacusSummit} suite provides multiple nested definitions of halos out of which we use the level two (L2) halos \citep[see,][for the details of the halo definition]{2022MNRAS.509..501H} We use center of mass position and velocity of the largest L2 subhalo fields, \texttt{x\underline{ }L2com} and \texttt{v\underline{ }L2com}, and generate our mocks in redshift space. \textsc{AbacusSummit} simulations come with a subsample (3 $\%$) of particles that make each halo, which will then be used for satellite population. Based on the HOD parameters we chose, galaxies with host halo mass lower than $10^{11} h^{-1}M_{\odot}$ barely exist. We then set a cut-off mass and remove all halo with mass smaller than $10^{11} h^{-1}M_{\odot}$ when populating HOD mock catalogs for all tracers (see App.\ref{app:downsamp}).

\begin{table}
\renewcommand{\arraystretch}{1.2}
\begin{tabularx}{\columnwidth}{|>{\setlength\hsize{1.1\hsize}\raggedright}X||>{\setlength\hsize{0.9\hsize}\centering}X|>{\setlength\hsize{0.9\hsize}\centering}X|>{\setlength\hsize{0.9\hsize}\centering}X|}
\cline{1-4}
 Parameters  & LRG & ELG & QSO \tabularnewline 
\cline{1-4}
$\log(M_{\rm cut})$ &12.70  &11.70  &12.50  \tabularnewline
$\sigma$            &0.17   &0.08   &0.30   \tabularnewline
$\log(M_{\rm 1})$   &13.80  &12.00  &15.00  \tabularnewline
$\log(M_{\rm 0})$   &12.13  &11.60  &12.00  \tabularnewline
$\alpha$            &1.28   &0.33   &1.20   \tabularnewline
%\cline{1-4}
$A_{\mathrm{c}}$               &0.70   &0.025  &0.05  \tabularnewline
$A_{\mathrm{s}}$               &0.70   &0.03   &1.00  \tabularnewline
%\cline{1-4}
$n$&5.14&6.35   &0.415   \tabularnewline
\cline{1-4}
$z$ &0.8  &1.1  &1.4  \tabularnewline
\cline{1-4}
\end{tabularx}
\caption{Fiducial values of HOD parameters for each tracer and the resulting comoving number density in units of $10^{-4}(\hMpc)^{-3}$.}
\label{tab:fidparams}
%\vspace{0.2in}
\end{table}

\begin{figure}
\centering
\includegraphics[width=0.9\linewidth]{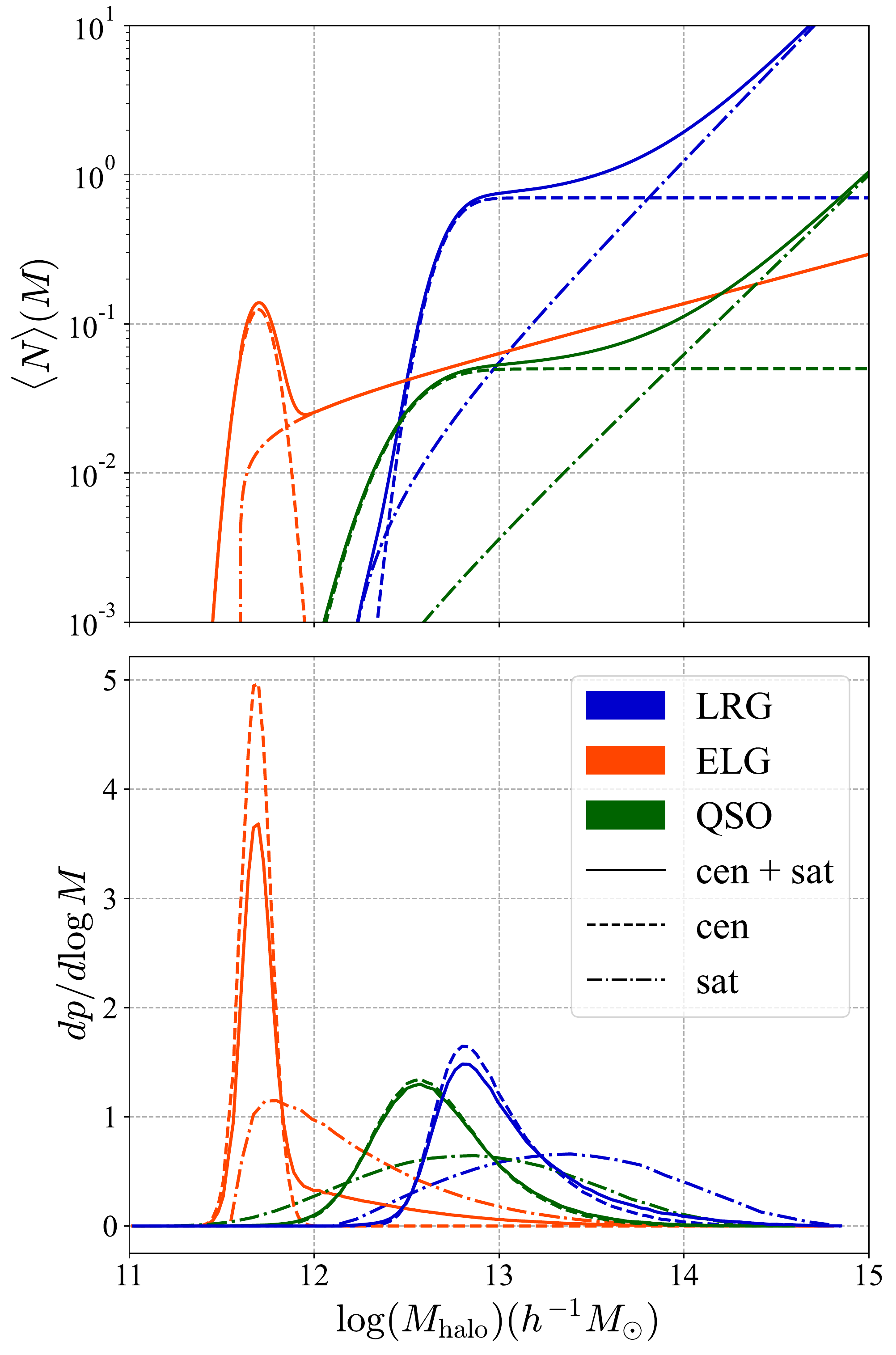}
\caption{\protect\footnotesize The top panel shows the expected number of galaxies hosted by a halo as a function of halo mass for the fiducial HOD parameters. The blue, orange, and green colors are for the LRG, ELG, and QSO respectively. The solid, dash and dash-dotted lines represent the expected number of all (cen+sat), central, and satellite galaxies. The bottom panel shows the probability distribution of host halo mass for a galaxy of each tracer. The solid line shows the host halo mass distribution for all, normalized as the probability per $\log(M)$, dash and dash-dotted line shows central and satellite host halo mass distribution respectively.}
\label{fig:hodmodel}
\end{figure}

Tab.~\ref{tab:fidparams} shows the fiducial parameter values that we use to create LRG, ELG, and QSO catalogs. They were obtained by fitting to the early version of the DESI Survey Validation data. These values may change as more DESI data is accumulated. For the purposes of our project, however, the exact fiducial values do not matter. The top panel on Fig.~\ref{fig:hodmodel} shows the expected number of galaxies per halo and as a function of the halo mass, the bottom panel shows the probability distribution of host halo mass for a galaxy normalized as the probability per $\log(M)$, based on fiducial parameter values in Tab.~\ref{tab:fidparams}. The host halo mass of ELG is smaller and more concentrated than that of LRG and QSO.

For each dark matter halo, we make a random decision whether to put a central galaxy in it and how many satellites (if any) we put in the halo. We compute a probability of a central galaxy and make a random draw from Bernoulli distribution $B(1,\langle N^c \rangle)$ and if it results in 1 we put a galaxy in the center of the halo. As for satellites, we chose the particle-based population approach \citep[e.g.][]{Yuan_2018}, we compute the average number of satellites then make a random draw from Bernoulli distribution $B(1,\langle N^s \rangle/N_p)$ and if it results in 1 we put a galaxy in the particle position, where $N_p$ is the number of particles attached to the halo. The total number of galaxy distribution is then Poissonian, $p(N|M)=\mathrm{Pois}(\langle N|M \rangle)$.

We assign the velocity of the halo to the central galaxy and the velocity of the particle to the satellite galaxy. Recent works \citep{2015MNRAS.446..578G,2015MNRAS.453.4368G} have shown that there is evidence for the velocity bias, the velocity of galaxies being systematically different from the velocity of the dark matter field at the same position. We ignore velocity bias in this work. This should not affect our results for the reasons outlined in the previous paragraph.

The procedure for creating mock catalogs is intrinsically stochastic. Depending on the outcomes of random draws we can get many different equivalent realizations of the galaxy population following the same HOD model on average.

\subsection{Projected correlation functions}

2PCF, $\xitwo(\mathbf{r)}$, describes a probability of finding two galaxies in infinitesimal volumes $dV_1$ and $dV_2$ that are separated by distance $\mathbf{r}$. This probability is proportional to $dV_1dV_2[1 + \xitwo\mathbf{r})]$ and is conventionally normalized so that for particles distributed in space with uniform probability (all spatial points are equally likely to host a particle) $\xitwo(\mathbf{r}) = 0$. The 3PCF, $\xithree(\mathbf{r}_{12},\mathbf{r}_{23},\mathbf{r}_{31})$, is similarly defined as an excess probability (over spatially uniform distribution) of finding a triplet of galaxies to be separated by $\mathbf{r}_{12}$, $\mathbf{r}_{23}$, and $\mathbf{r}_{31}$ \citep{1980lssu.book.....P}.  2PCF of observed galaxies depends only on the along and across the line-of-sight separations (with respect to the observer) of galaxies instead of the full separation vector, $\xitwo(\mathbf{r)} = \xitwo(r_p, \pi)$ where $r_p$ is a distance perpendicular to the line-of-sight and $\pi$ is a distance along the line-of-sight. The 3PCF similarly depends on three perpendicular separations and two relative distances along the line of sight, $\xithree (\mathbf{r}_{12},\mathbf{r}_{23},\mathbf{r}_{31}) = \xithree (r_{p12},r_{p23}, r_{p31}, \pi_{12}, \pi_{23})$. The variations in the line-of-sight separation in these correlation functions depend on the velocities of the galaxies in addition to their positions. To make HOD modeling easier projected correlation functions are often used \citep{1983ApJ...267..465D,2004ApJ...614..527Z}. They are defined by
\begin{align}
    \wptwo(r_{\mathrm{p}})&=\displaystyle\int\limits_{-\pi^\star}^{\pi^\star}\!\!\!\!\! d\pi \xitwo(r_{\mathrm{p}},\pi)\label{eq:wptwoint}\\ 
    \wpthree(r_{\mathrm{p12}},r_{\mathrm{p23}},r_{\mathrm{p31}})&=\displaystyle\int\limits_{-\pi^{\star}}^{\pi^{\star}}\!\!\!\!\!d\pi_{1}d\pi_{2}
 \xithree(r_{\mathrm{p12}},r_{\mathrm{p23}},r_{\mathrm{p31}},\pi_{1},\pi_{2})\label{eq:wpthreeint}
\end{align}

The value of $\pi^\star$ is usually chosen to be of the order of a few tens of megaparsecs. This is done to smooth over peculiar velocity effects that affect the functional dependence of the correlation functions in the parallel to the line-of-sight direction. We derive our main results using the value of $\pi^\star = 100h^{-1}$ Mpc for the projected 2PCF. This value is large enough for the residual peculiar velocity effects to be negligible. Although, these projected correlation functions will depend on the velocities of the galaxies unless $\pi^\star \rightarrow \infty$ \citep[see e.g.][]{2009MNRAS.396...19N,2013MNRAS.430..725V}. Lower values of $\pi^\star$ maybe optimal because they do not depend on large scale correlations that are noisier, but using a lower integration limit would require careful modelling of the peculiar velocity effects and the difference turns out not to be big enough to affect any of our main conclusions (see App.~\ref{app:pistar}).

\subsection{Measuring projected correlation functions}

\begin{figure}
\centering
\includegraphics[width=0.9\linewidth]{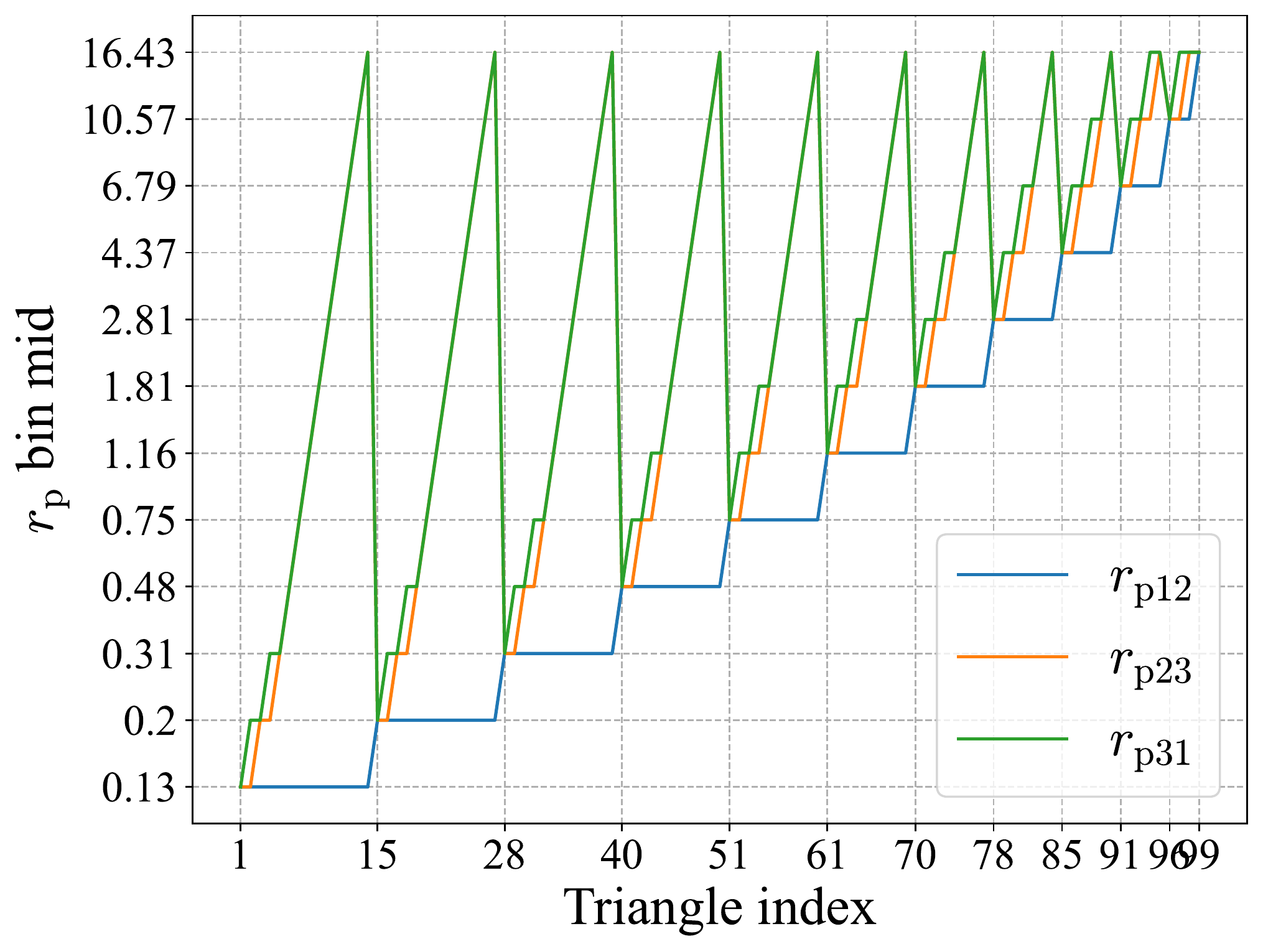}
\caption{\protect\footnotesize projected separation as a function of the triangular index.}
\label{fig:triind}
\end{figure}

\label{sec:covmat}
\begin{figure*}
\centering
\subfigure {\includegraphics[width=0.33\textwidth]{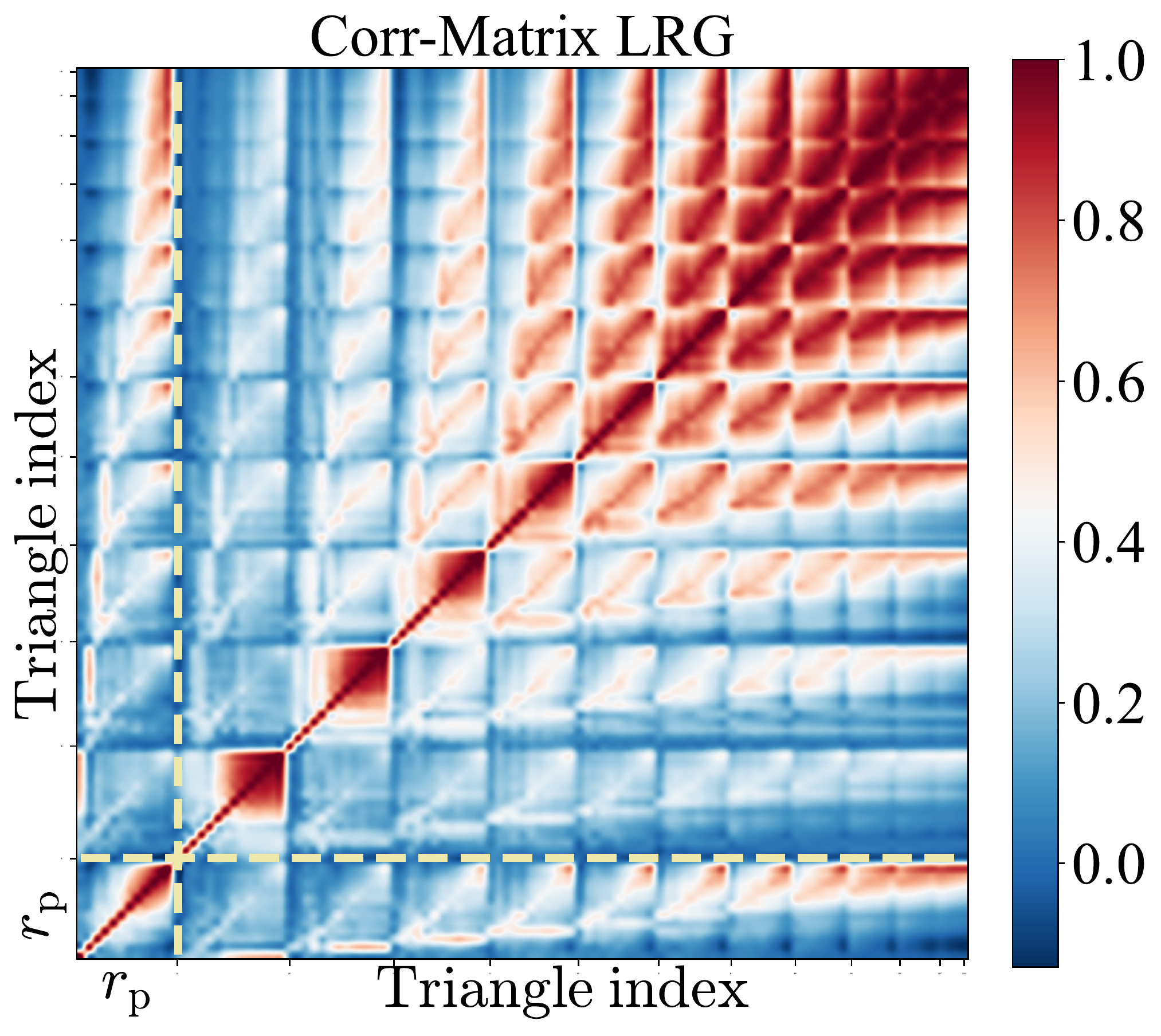}}
\subfigure {\includegraphics[width=0.33\textwidth]{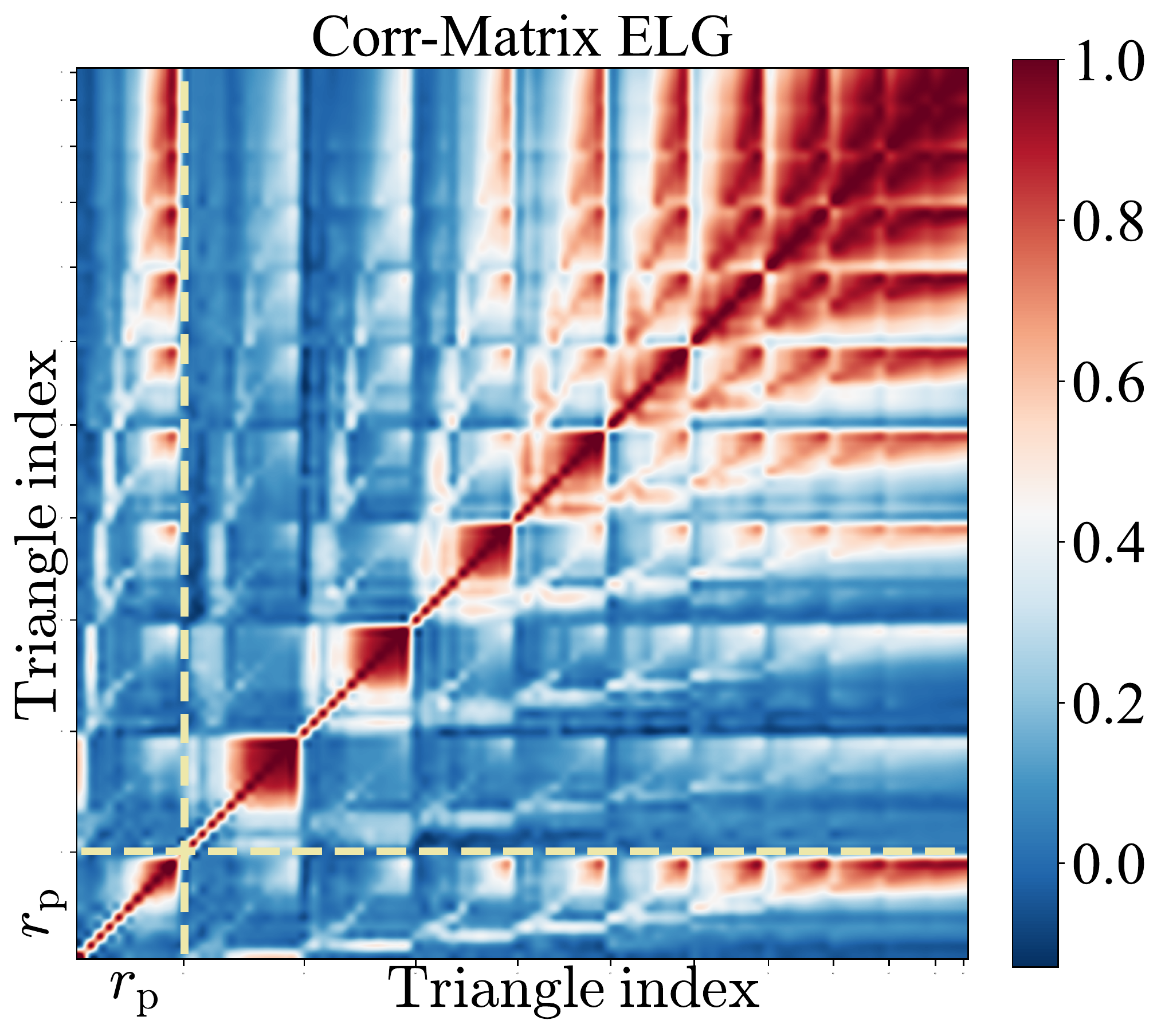}}
\subfigure {\includegraphics[width=0.33\textwidth]{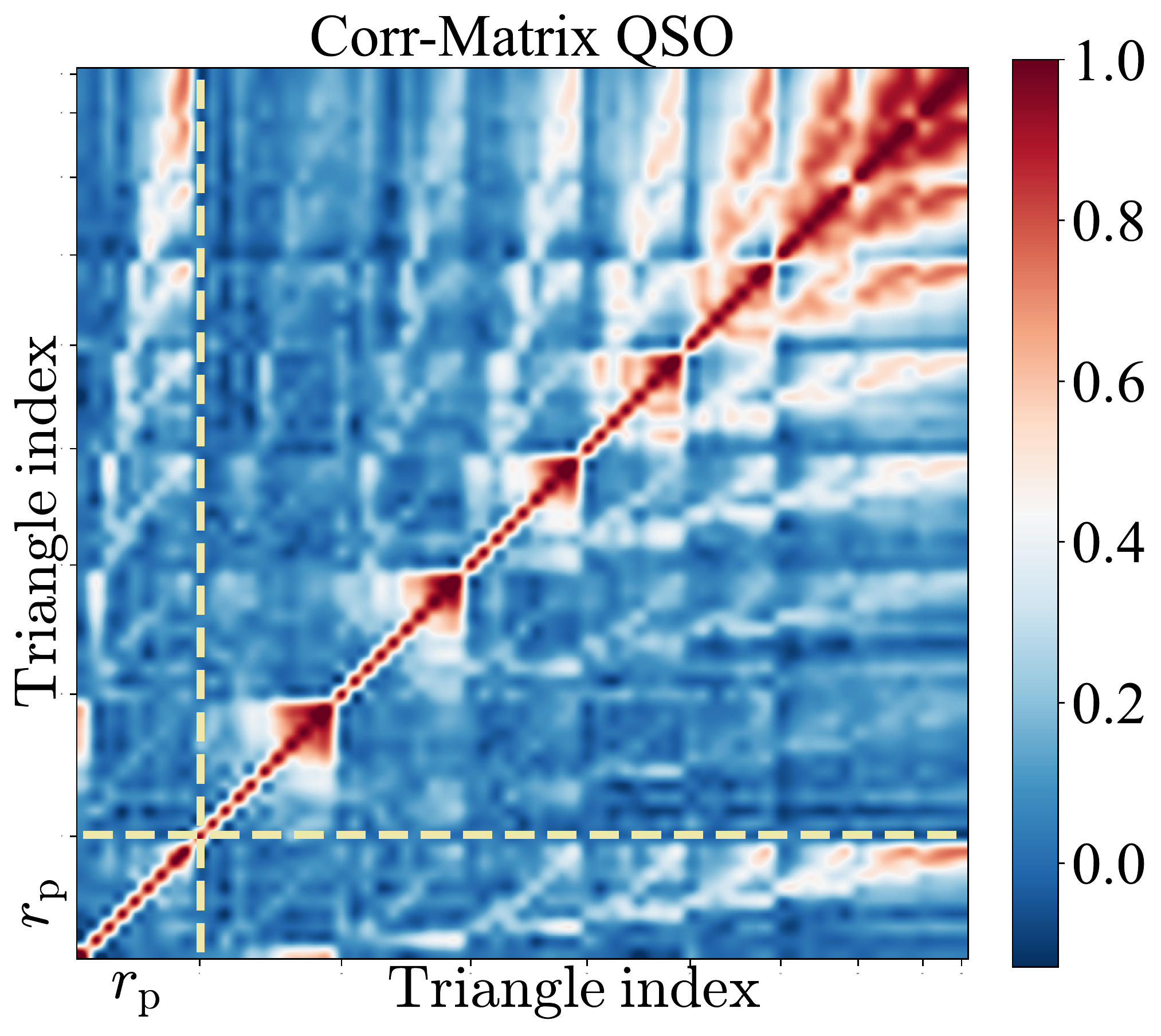}}
\caption{\protect\footnotesize Correlation matrices including the projected 2PCF and projected 3PCF (simplified version) for LRGs (left), ELGs(mid), and QSOs(right) used in this analysis derived by Jackknife re-sampling. We include all bins between $r_{\mathrm{p}}=0.1 - 20 \hMpc$ for the projected 2PCF, all triangles of LRGs, triangle index from 15 to 99 of ELGs, and triangle index from 28 to 99 of QSOs for the projected 3PCF. The color indicates the level of correlation, where red represents $100\%$ correlation and dark blue means a low level of anti-correlation.}
\label{fig:corr}
\end{figure*}

The 2PCF and 3PCF are usually measured by counting the number of galaxy pairs and triplets for the data and for uniform distribution in the same volume. They can be estimated from these pair and triplet counts by
\begin{align}
    {\xi}^{(2)}(r_{\mathrm{p}},\pi) &= \frac{DD(r_{\mathrm{p}},\pi)}{RR(r_{\mathrm{p}},\pi)} - 1, \label{eq:xi2dd}\\ 
    {\xi}^{(3)}(r_{\mathrm{p12}},r_{\mathrm{p23}},r_{\mathrm{p31}},\pi_{1},\pi_{2}) &= \frac{DDD(r_{\mathrm{p12}},r_{\mathrm{p23}},r_{\mathrm{p31}},\pi_{1},\pi_{2})}
    {RRR(r_{\mathrm{p12}},r_{\mathrm{p23}},r_{\mathrm{p31}},\pi_{1},\pi_{2})} - 1, 
    \label{eq:xi3ddd}
\end{align}
where $DD$ is the number of pairs of galaxies separated by certain radial and transverse distances, $DDD$ is the number of triplets of galaxies having a specific triangular configuration, $RR$ and $RRR$ are the equivalent number of pairs and triplets from a uniform random distribution. Additive factors of -1 normalize the correlations to be zero when $DD \sim RR$ and $DDD \sim RRR$. 

We compute the 2-point projected correlation functions by estimating the 2PCF first and then integrating the estimated correlation functions over $\pi$. For the 3-point projected correlation function, we use a slightly modified algorithm. Instead of eq.\eqref{eq:wpthreeint}, we compute a simplified version (SV),
\begin{align}
        w^{(3)}_{\mathrm{p(SV)}} &= \frac{\displaystyle\sum_{\pi_1,\pi_2} DDD(r_{\mathrm{p12}},r_{\mathrm{p23}},r_{\mathrm{p31}},\pi_{1},\pi_{2})}{\displaystyle\sum_{\pi_1,\pi2} RRR(r_{\mathrm{p12}},r_{\mathrm{p23}},r_{\mathrm{p31}},\pi_{1},\pi_{2})}-1 \nonumber\\
        &= \frac{DDD(r_{\mathrm{p12}},r_{\mathrm{p23}},r_{\mathrm{p31}})}
    {RRR(r_{\mathrm{p12}},r_{\mathrm{p23}},r_{\mathrm{p31}})} - 1. \label{eq:wp3SV}
\end{align}
This is not the same projected correlation function as the one defined in eq.~\eqref{eq:wpthreeint}. Eq.\eqref{eq:wpthreeint} compute the integral over $\pi_1, \pi_2$, which still need triangle counting in a 5 dimension space. Eq.~\eqref{eq:wp3SV}, on the other hand, compute a ratio of sums over $\pi_1, \pi_2$, which reduced triangle counting to a 3 dimension space only relay on $r_{\mathrm{p12}},r_{\mathrm{p23}},r_{\mathrm{p31}}$. What we estimate with eq.~\eqref{eq:wp3SV} is still a three-point function that depends on the distribution of triangular configurations and it is projected in a sense that it is insensitive to the radial separation between the three galaxies (and therefore also insensitive to the velocities of the galaxies). The second function is significantly easier and faster to compute. We choose the $\hat{z}$-direction of the \textsc{AbacusSummit} boxes to be the line-of-sight of the observer. This makes the projected distance along z the $\pi$ and the projected distance in the x-y plane the $r_\mathrm{p}$. This lets us completely ignore the $\hat{z}$-direction and significantly accelerate the triplet counting part of the algorithm. We use a modified version of \textsc{GANPCF} package \footnote{\url{https://github.com/dpearson1983/ganpcf}}, which is a GPU accelerated tool for N-point correlation function measurements, to compute the $DDD$ counts defined this way.

We measure the $DD$ counts of the projected 2PCF using \textsc{CorrFunc} package \citep{2020MNRAS.491.3022S,10.1007/978-981-13-7729-7_1} setting $\pi^\star = 100\hMpc$. A smaller value of $\pi^\star$ would result in a less noisy measurement, but since our main objective is to compare the relative constraining power of the 2 and 3 point clustering, we need to compute both in similar settings.

Our volume is a simple periodic cube and the $RR$ and $RRR$ counts for the uniform distribution can be computed analytically (see App.~\ref{app:anarand} for analytical $RRR$ computation).

The correlation functions change more rapidly at small separations. We require narrower bins at smaller separations in order not to lose too much information to the binning effects. To achieve this we measure $\wptwo(r_\mathrm{p})$ in 12 bins equally spaced in $\log_{10}r_\mathrm{p}$ between 0.1 $\hMpc$ and 20 $\hMpc$.

We use the same binning for the three sides of the $w_{\mathrm{p(SV)}}^{(3)}(r_{\mathrm{p12}},r_{\mathrm{p23}},r_{\mathrm{p31}})$. We arrange triplets of separations by starting with all possible unique triplets that satisfy $r_\mathrm{p12} \leq r_\mathrm{p23} \leq r_\mathrm{p31}$. We start with the triplet that has all three sides belonging to the shortest separation bin. We then arrange all other triplets so that each following triplet is in increasing order of $r_\mathrm{p12}$. Triplets that have equal $r_\mathrm{p12}$ we internally arrange by increasing $r_\mathrm{p23}$. Finally, the triplets that have both $r_\mathrm{p12}$ and $r_\mathrm{p23}$ equal we arrange by increasing $r_\mathrm{p31}$. We remove triplets for which the midpoints of the bins do not satisfy the triangular condition $r_{\mathrm{p31}}\leq r_{\mathrm{p12}}+r_{\mathrm{p23}}$. Once the triplets are arranged and sorted in this way, we assign to each one of them an integer ``triangular index''. For our choice of binning, we end up with 99 unique triangular configurations. Fig.~\ref{fig:triind} shows the values of $r_{\mathrm{p12}}$, $r_{\mathrm{p23}}$, $r_{\mathrm{p31}}$ as a function of the triangular index.

\subsection{Covariance matrix of projected correlation functions}

We use the jackknife re-sampling method to estimate the variance of clustering. We divide the simulation volume into $N_{\mathrm{sub}}$ sub-volumes. We compute the projected 2PCF and 3PCF by omitting each one of the subvolumes. This results in $N_\mathrm{sub}$ measurements corresponding to $(N_{\mathrm{sub}}-1)/N_{\mathrm{sub}}$ fraction of the origin volume. Covariance matrix from jackknife method is then estimated by:
\begin{equation} \label{eqn:jkcov}
    \begin{aligned}
        C_{i,j}^{\mathrm{jk}}= \frac{(N_{\mathrm{sub}}-1)}{N_{\mathrm{sub}}} 
        \sum_{k=1}^{N_{\mathrm{sub}}}(X_i^k-\Bar{X}_i)(X_j^k-\Bar{X}_j)
    \end{aligned}
\end{equation}
where $X_i^k$ is the clustering measurements (either 2PCF or 3PCF) in $i\mathrm{th}$ bin from the $k\mathrm{th}$ jackknife realization. Overline denotes an average measurement over all realizations.
\begin{equation} \label{eqn:jkmean}
    \begin{aligned}
        \Bar{X}_i=\frac{1}{N_{\mathrm{sub}}}\sum_{k=1}^{N_{\mathrm{sub}}}X_i^k.
    \end{aligned}
\end{equation}
This version of the jackknife realization is referred to as ``delete-one'' version in the literature.

We set $N_{\mathrm{sub}}=400$ to make sure we have enough sub-volumes to estimate the error of 3PCF. We slice the box $Z$ (LOS) axis into rectangles with equal area squared base on $XY$ plane. 
Fig.~\ref{fig:corr} shows the correlation matrices for LRGs, ELGs and QSOs measured from the jackknife method using HOD mock catalog we populated as described in Sec.~\ref{sec:mock}. Each panel shows a matrix with horizontal and vertical division dash lines. The first column displays the correlation between $r_{\mathrm{p}}$ bins in the projected 2PCF with itself (bottom), and with the triangles in the projected 3PCF(top). The second column is the correlation between the 2PCF and 3PCF (bottom) and 3PCF with itself. For the correlation matrix of LRGs, there is a strong auto-correlation for the projected 2PCF at scale $r_{\mathrm{p}}>1.4\hMpc$ (after 6th bin) and for the projected 3PCF with at least two triangle side lengths $r_{\mathrm{p}}>1.4\hMpc$. There is a cross-correlation between 2PCF at relatively large scale and 3PCF with at least two triangle side lengths at corresponding scale, while other cross-correlation is quite weak. The correlation matrices of ELGs and QSOs show similar pattern as LRGs, with a weaker correlation.

These covariances correspond to the constraining power of a one cubic Gigaparsec box. The actual DESI samples will cover a much larger volume. For small separations the covariances on both the 2PCF and the 3PCF will scale as an inverse of a volume.

\begin{figure*}
\centering
\includegraphics[width=\linewidth]{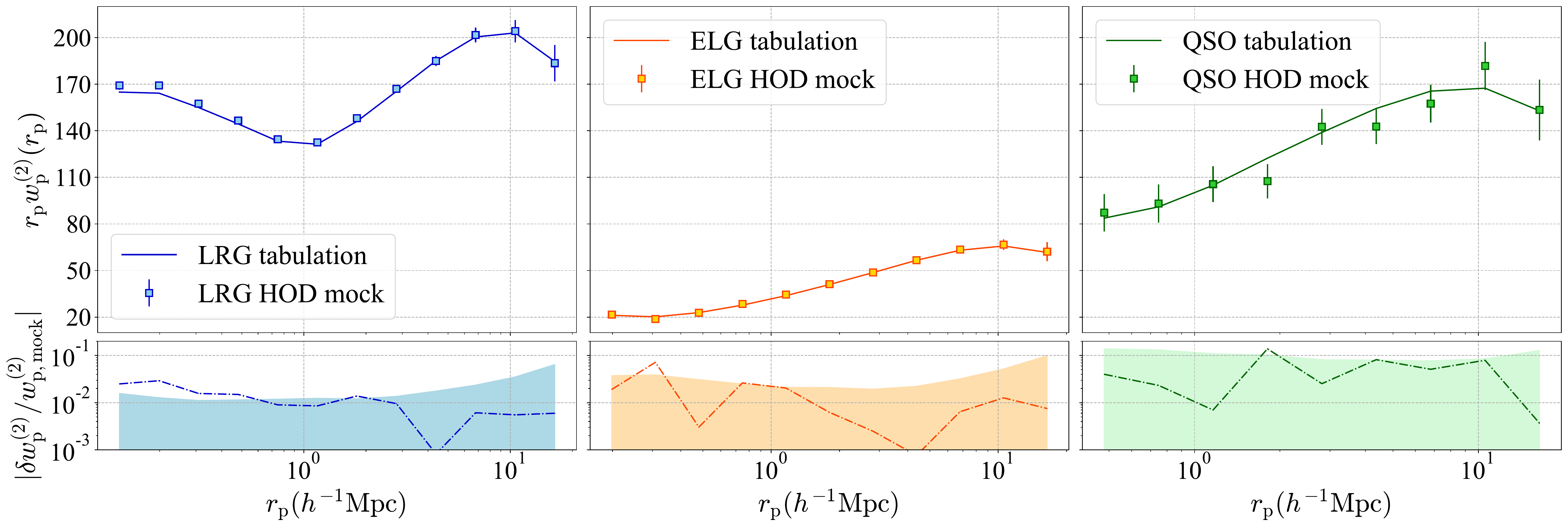}
\caption{\protect\footnotesize Panels on top are the projected 2PCF from HOD mock catalog and tabulation method with fiducial HOD parameters for different tracers. The blue, orange and green colors are for LRGs, ELGs, and QSOs respectively. Filled markers are measurements from the fiducial HOD mock catalog, errors are calculated using the jackknife method. Solid lines represent measurements from the tabulation method. Dot-dash lines on the bottom sub-panel are absolute value of percentage difference the light shaded area represents jackknife error.}
\label{fig:wp2}
\end{figure*}

\begin{figure*}
\centering
\includegraphics[width=1\linewidth]{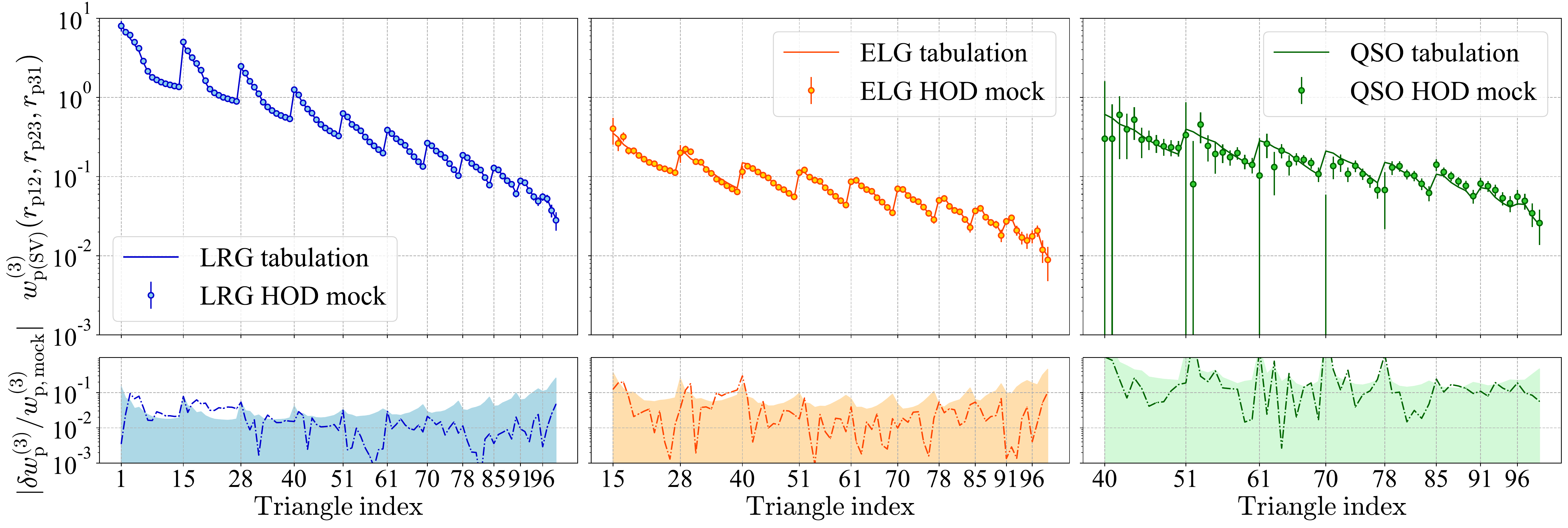}
\caption{\protect\footnotesize A similar plot for 3PCF. Panels on top are the simplified projected 3PCF from HOD mock catalog and tabulation method with fiducial HOD parameters for different tracers. The blue, orange, and green colors are for LRGs, ELGs, and QSOs respectively. Filled markers are measurements from the fiducial HOD mock catalog, errors are calculated using the jackknife method. Solid lines represent measurements from the tabulation method. Dot-dash lines on the bottom sub-panel are absolute value of percentage difference the light shaded area represent jackknife error.}
\label{fig:wp3}
\end{figure*}

Solid circles in Fig.~\ref{fig:wp2} show the projected 2PCF measurements from the mock catalog for different tracer and the jackknife errorbars. The first bin of the ELG and the first three bins of the QSO projected correlation function has been omitted. For ELGs, it is a conservative choice to omit the smallest scale bin (see App.\ref{app:downsamp}). For QSOs, the number density of QSOs is too small to have a sufficient number of pairs on those small scales. Fig.~\ref{fig:wp3} shows a similar plot for the projected 3PCF where all triangles that include the bins omitted for the 2PCF have been removed. This results in 99, 85, 60 triangular configurations for LRGs, ELGs, and QSOs respectively. We keep the original triangular indexes that have been assigned before the removal of the low separation triangles. As a result, the ELG triangular index starts with 15, and the QSO triangular index starts with 40.

\subsection{Constraining HOD parameters}
\label{sec:mcmc}

Not all three galaxy samples we consider can be constrained equally well with data from a one cubic Gigaparsec box. We find that for the LRGs it is possible to constrain all five parameters: $\log(M_{\rm cut})$, $\sigma$, $\log(M_{\rm 1})$, $\log(M_{\rm 0})$ and $\alpha$ ($A_{\mathrm{c}}$ and $A_{\mathrm{s}}$ are used for the tuning of the number density of the LRG sample and do not affect the 2 and 3 PCF). For ELGs, constraining all five parameters turns out to be more difficult. We only let the $\log(M_{\rm cut})$, $\alpha$ and $A_{\mathrm{s}}$ be free parameters and fix the remaining two to their fiducial values. $A_{\mathrm{s}}$ is degenerate in its effects with $\log(M_{\rm 1})$. We choose to vary $A_{\mathrm{s}}$ in our computations for convenience. For QSOs, we need to further reduce the number of free parameters because the QSO sample has a much lower number density. We set $\log(M_{\rm cut})$ and $\log(M_{\rm 1})$ as free parameters and fix the remaining three to their fiducial values. We apply flat priors for all free parameters. The intervals are listed in Tab.~\ref{tab:prior}. The fiducial values for the fixed parameters are listed in Tab.~\ref{tab:fidparams}.

We perform Markov Chain Monte Carlo (MCMC) to obtain the posterior probability distribution of parameter space. The likelihood function $\mathcal{L} \propto \exp{-\chi^2/2}$, where $\chi^2$ is given by

\begin{equation}
    \chi^2=\Delta X_i (S^\prime)^{-1}_{ij} \Delta X_j,
\end{equation}

where $\Delta X_i$ is the difference of binned 2PCF and 3PCF between theory and observation, in our case corresponding to tabulated estimation and HOD mock measurements. $(S^\prime)^{-1}$ is the inverse of the re-scaled covariance matrix, here we follow \citet{percival2021matching} to take into account the error propagation from the error in the covariance matrix into the fitting parameters. 

\begin{align}
    &S^{\prime}=\frac{(n_s-1)[1+B(n_d-n_p)]}{n_s-n_d+n_p-1}S \\
    &B=\frac{(n_s-n_d-2)}{(n_s-n_d-1)(n_s-n_d-4)}
\end{align}

Where $n_s$ is the number of jackknife realizations, $n_d$ is the number of data points we are fitting to and $n_p$ is the number of free parameters in the model. $S$ is the original covariance matrix.

We use a modified version of \textsc{CosmoMC} \citep{Lewis:2002ah} as MCMC engine ,which use the Metropolis-Hastings algorithm, to sample the parameter space and search for the minimum $\chi^2$. We run 16 chains parallel with MPI and we ignore the first $30$\% of each chain as burn-in. We apply the Gelman and Rubin R statistic \citep{An98stephenbrooks} as convergence criteria, all of our chains have $R-1 < 0.01$, which represents a good convergence. Since the tabulation method provide us a fast estimation of projected 2PCF and 3PCF (see Sec.\ref{sec:tabcorr}), the MCMC stage takes less than half an hour to converge for a joint run using both 2PCF and 3PCF. 
\begin{table}
\centering
\renewcommand{\arraystretch}{1.2}
\begin{tabularx}{\columnwidth}{|>{\setlength\hsize{0.8\hsize}\raggedright}X||>{\setlength\hsize{0.9\hsize}\centering}X|>{\setlength\hsize{0.9\hsize}\centering}X|>{\setlength\hsize{0.9\hsize}\centering}X|}
\cline{1-4}
 Parameters  & LRG & ELG & QSO \tabularnewline 
\cline{1-4}
$\log(M_{\rm cut})$ &[12.0, 13.5]   &[11.0, 13.5]   &[11.5, 13.5]  \tabularnewline
$\sigma$            &[0.0001, 1.0]  &-              &-   \tabularnewline
$\log(M_{\rm 1})$   &[12.5, 14.5]   &-              &[13.0, 17.0]  \tabularnewline
$\log(M_{\rm 0})$   &[11.0, 15]     &-              &-  \tabularnewline
$\alpha$            &[0.0, 2.0]     &[0.0, 2.0]     &-   \tabularnewline
$A_{\mathrm{c}}$    &-              &-              &-  \tabularnewline
$A_{\mathrm{s}}$    &-              &[0.0, 0.2]     &-  \tabularnewline
\cline{1-4}
\end{tabularx}
\caption{The flat prior interval on HOD parameter for different tracer, fitting to LRG, ELG, and QSO HOD mock catalog have 5, 3, and 2 free parameters respectively. Parameters with a dash are fixed to their fiducial values. }
\label{tab:prior}
%\vspace{0.2in}
\end{table}

\section{Tabulation method of computing galaxy N-point Correlation Functions}
\label{sec:tabcorr}

\begin{figure*}
\centering
\includegraphics[width=\linewidth]{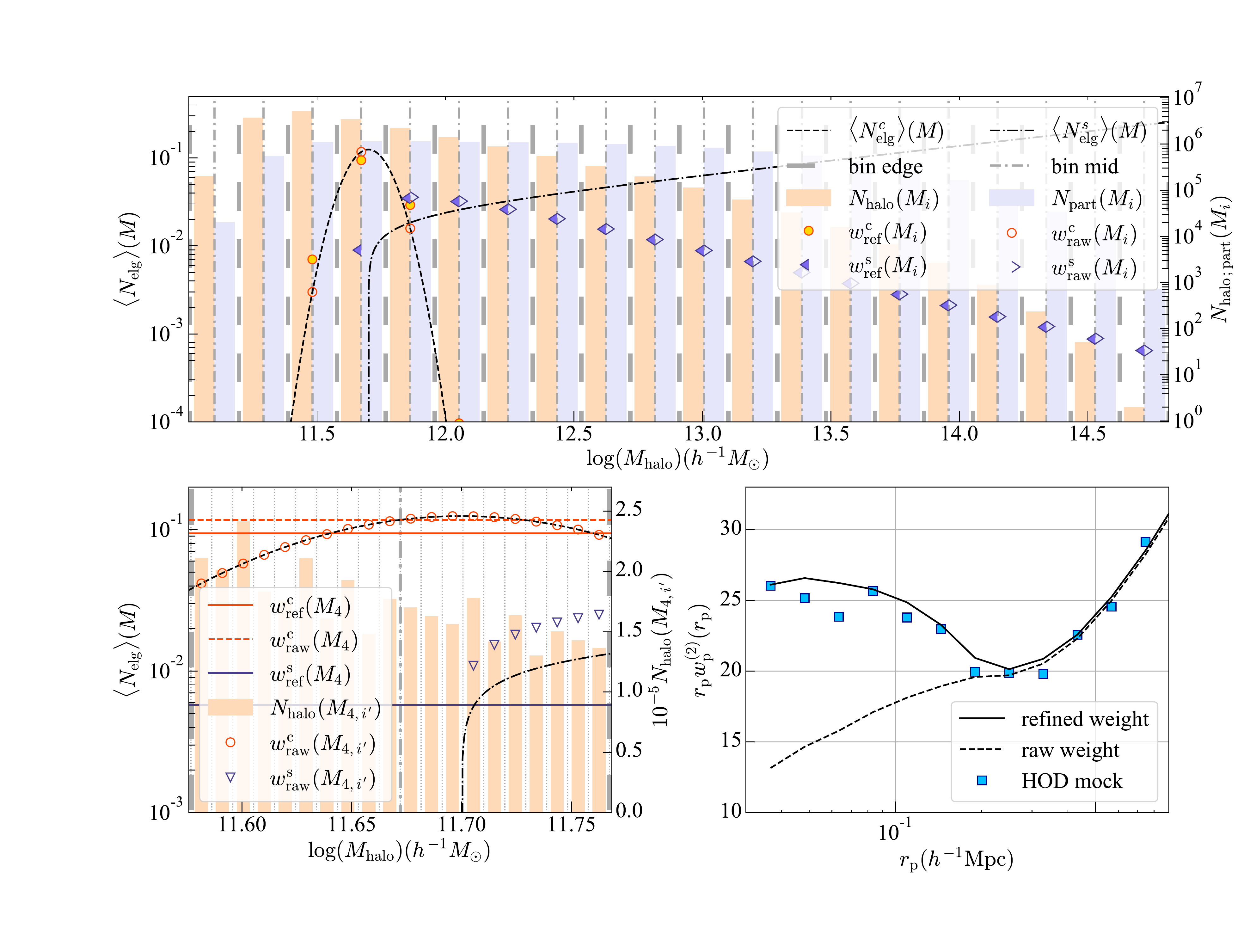}
\caption{\protect\footnotesize Mass binning effects for ELGs when shift satellite parameter $\log(M_0)$ slightly to $11.7$ from fiducial, all other parameters remain at fiducial value.} Top panel: Gray vertical lines show the edges (dashed) and the middle points (dashed-dotted) of the mass bins. The lines show the expected number of centrals (dashed) and satellites (dashed-dotted). The bars show the available number of halos (peach) and particles (lavender) in the simulation. Triangles show the weight of satellites computed with the raw probabilities (open) and refined probabilities (filled). The open and filled circles show the same information for the centrals. Bottom left panel: A zoom-in version of the $4$th mass bin, where raw weights do not work. Thin dotted vertical lines show the edges of sub mass bin. The peach bars show the number of halos in each sub-mass bin. The open circles and triangles show values of raw weights for each sub mass bin. The orange horizontal dash line shows raw central weight for $4$th mass bin and the orange solid line shows refined central weight for $4$th mass bin. The solid purple line shows refined satellite weight for $4$th mass bin and there is no dash purple line for raw satellite weight because the value is zero. Bottom right panel: Blue squares show the measured projected 2PCF of the galaxies from the ELG HOD mock catalog using HOD parameters plotted. Lines show the projected 2PCF computed with the tabulated method using the raw (dashed) and refined (solid) weights.
\label{fig:tab}
\end{figure*}

\subsection{Tabulated 2PCF}

The two steps leading to the pair counts of the mock catalog --- populating $N$-body mocks with galaxies and counting pairs of galaxies --- can be formally summarized with the equation
\begin{equation}\label{eqn:DD_rp}
    DD(r_p) = \displaystyle\sum_{ij}\Theta^{r_p}_{ij}D_iD_j,
\end{equation}
where each index in the double summation goes over all halo centers and halo particles,
\begin{equation}
    \Theta^{r_p}_{ij} = 
    \begin{cases}
        1, & \text{if distance between (i,j) pair falls within} \\
        & \text{the specified $r_p$ bin}\\
        0, & \text{otherwise}
    \end{cases}
\end{equation}
and $D$ is a stochastic variable
\begin{equation}
    D_i = 
    \begin{cases}
        1, & \text{if the $i^\mathrm{th}$ halo/particle got populated by a galaxy}\\
        0, & \text{otherwise}
    \end{cases}
\end{equation}
We rewrite eq.~\eqref{eqn:DD_rp} as
\begin{align}
    DD&= \displaystyle\sum_{ij}^{N_h}\Theta^{hh}_{ij}D^h_iD^h_j \\
    &+2\displaystyle\sum_{i,j}^{N_h,N_p}\Theta^{hp}_{ij}D^h_iD^p_j + \displaystyle\sum_{ij}^{N_p}\Theta^{pp}_{ij}D^p_iD^p_j \nonumber
\end{align}
explicitly separating halos and particles, where superscripts h and p refer to the halos and particles respectively, $N_h$ and $N_p$ are the numbers of halos and particles, and we dropped the $r_p$ label for brevity. In the traditional approach (top panel of Fig.~\ref{fig:flowchart}) the random numbers $D$ have to be drawn for and the double sum over all occupied halos and particles computed for every HOD model under consideration.

The tabulation approach reduces the complexity of this computation by employing the following trick. The expectation value of the pair count is
\begin{align}\label{eqn:DD_tran}
    \langle DD \rangle = \displaystyle\sum_{ij}^{N_h}\Theta^{hh}_{ij} \lambda^c_i \lambda^c_j + 2\displaystyle\sum_{i,j}^{N_h,N_p}\Theta^{hp}_{ij}\lambda^c_i \lambda^s_j + \displaystyle\sum_{ij}^{N_p}\Theta^{pp}_{ij} \lambda^s_i \lambda^s_j,
\end{align}
where $\lambda^c$ and $\lambda^s$ are the expected values of that particular halo or a particle to host a central or satellite galaxy in a given HOD model. Since these numbers only depend on the mass of the host halo, we can simplify the computation by binning the halos and particles into bins of mass in log space narrow enough so that the expected values do not significantly change within it. The pair count can then be rewritten as
\begin{align}
        \langle DD \rangle &= \displaystyle\sum_{i,j}^{N_h^k,N_h^\ell}\displaystyle\sum_{k\ell}^{N_b}\Theta^{hh}_{ij,k\ell} \bar{\lambda^c_k} \bar{\lambda^c_\ell} \\
        &+ 2\displaystyle\sum_{i,j}^{N_h^k,N_p^\ell}\displaystyle\sum_{k\ell}^{N_b}\Theta^{hp}_{ij,k\ell} \bar{\lambda^c_k} \bar{\lambda^s_\ell} +
        \displaystyle\sum_{i,j}^{N_p^k,N_p^\ell}\displaystyle\sum_{k\ell}^{N_b}\Theta^{pp}_{ij,k\ell} \bar{\lambda^s_k} \bar{\lambda^s_\ell},\nonumber\\
\end{align}

where the indices $k$ and $\ell$ now go over $N_b$ number of mass bins, $N_h^k$ is the number of halos in the $k^\mathrm{th}$ mass bin (similarly for particles) and $\bar{\lambda^c}$ and $\bar{\lambda^s}$ are effective average expected values in each mass bin, 

\begin{align}
        \bar{\lambda^c_k} &= \langle N^{c}\rangle(M_k)\\
        \bar{\lambda^s_k} &= \langle N^{s}\rangle(M_k)\frac{N_h^k}{N_p^k}
\end{align}

$M_k$ is the representative mass in $k$th mass bin $\log{M_k}\pm (\Delta\log M)/2$, where $\Delta\log M$ is the width of log space mass bin. $\langle N^{c}\rangle(M_k)$ and $\langle N^{s}\rangle(M_k)$ are expectation number of central and satellite galaxies hosted by halo in $k$th mass bin. $\bar{\lambda^s_k}$ could be understood this way, $\langle N^{s}\rangle(M_k)N_h^k$ is the expected total amount of satellite galaxies in $k$th mass bin, divided by the $N_p^k$ would give us the average expected values for each particle to host a satellite galaxy.

Switching the order of summation we get
\begin{align}\label{eqn:DD_tabcorr}
        \langle DD \rangle &= \displaystyle\sum_{k\ell}^{N_b} \bar{\lambda^c_k} \bar{\lambda^c_\ell} DD^{hh}_{k\ell}\\
        &+2\displaystyle\sum_{k\ell}^{N_b} \bar{\lambda^c_k} \bar{\lambda^s_\ell} DD^{hp}_{k\ell}
        +\displaystyle\sum_{k\ell}^{N_b} \bar{\lambda^s_k} \bar{\lambda^s_\ell} DD^{pp}_{k\ell},\nonumber
\end{align}
where
\begin{equation}
    DD^{hh}_{k\ell} = \displaystyle\sum_{i,j}^{N_p^k,N_p^\ell}\Theta^{hh}_{ij,k\ell}
\end{equation}
is the number of halo pairs with a separation that falls in the specified bin, where one member of the pair is in mass bin $k$ while another is in mass bin $\ell$ (similarly for the particle-halo and particle-particle pairs).

Eq.~\eqref{eqn:DD_tabcorr} is equivalent to the eq.~\eqref{eqn:DD_tran} but has two advantages. First, it gives an average value of the number count expected for a given HOD model instead of a specific realization that includes stochastic noise. Secondly, it has the potential to save a significant amount of computational time. The most time-consuming part of the computation - the double sum over halos and particles - can be performed only once. Changing the HOD model amounts to simply summing up precomputed pair counts with different weights, a procedure that is orders of magnitude faster.

This method was used by \citep{2016MNRAS.458.4015Z} to estimate 2PCF from N-body simulations efficiently. The approach we introduced above is more like section 2.2 in \citet[Case with Subhaloes]{2016MNRAS.458.4015Z}, instead of subhaloes, we populate satellite with particles here. This method could be applied to different kinds of galaxy clustering, e.g. real space 2PCF, projected 2PCF, and 2PCF multipole. The correlation function is given by a similar weighted sum over different mass bin cross-correlations and we take the projected correlation function as an example to show in detail.

Halos and particles live in the same periodic box so the $RR$ counts are identical for them. This means that the projected 2PCF is also a weighted average of the cross-2PCF of different mass halos (and particles)
\begin{equation} \label{eqn:xi2tab}
    \begin{aligned}
        w^{(2)}_{\mathrm {p,gg}}(r_\mathrm{p})&=\sum_{k,\ell}^{N_b}\wcen(M_k)\wcen(M_\ell)w^{(2)}_{\mathrm{p,hh}}(r_\mathrm{p},M_k,M_\ell)\\
                &+2\sum_{k,\ell}^{N_b} \wcen(M_k)\wsat(M_\ell)w^{(2)}_{\mathrm{p,hp}}(r_\mathrm{p},M_k,M_\ell)\\
                &+\sum_{k,\ell}^{N_b} \wsat(M_k)\wsat(M_\ell)w^{(2)}_{\mathrm{p,pp}}(r_\mathrm{p},M_k,M_\ell),
    \end{aligned}
\end{equation}
where $w^{(2)}_{\mathrm{p,hh}}(r_\mathrm{p},M_i,M_j)$ is the two-point cross correlation function of halos in the $i$th and $j$th mass bins (similarly for the halo-particle and particle-particle correlation functions) and naively we could take the weight as
\begin{align}
    \wcen_{\mathrm{raw}}(M_k) &= \bar{\lambda^c_k} = \langle N^{c}\rangle(M_k)\label{eqn:wcraw},\\
    \wsat_{\mathrm{raw}}(M_k) &= \bar{\lambda^s_k} = \langle N^{s}\rangle(M_k)\frac{N_h^k}{N_p^k}\label{eqn:wsraw}.
\end{align}

We define Eqn. \eqref{eqn:wcraw}$\And$\eqref{eqn:wsraw} as raw weights, raw weights are a good approximation in most cases but have some exceptions, we will further explain this in Sec.~\ref{sec:massbineffect}.

Solid lines on the top panels of Fig.~\ref{fig:wp2} show the galaxy projected 2PCF computed using the tabulation method. The bottom panels show the fractional deviation between the projected 2PCF computed with the tabulated method and a specific realization. The offset is in all cases within the expected standard deviation. The deviations are caused by the stochasticity in the realization, the tabulated method being almost noise-free. There is a very small stochastic noise in the tabulated 2PCF related to the finite number of halos and particles in the box, but it is negligible compared to the noise in a single realization.

\subsection{Tabulated 3PCF}

We further generalize derivation in the previous section to the 3PCF. Similar arguments lead to the expression
\begin{equation} \label{eqn:xi3tab}
    \begin{aligned}
        w^{(3)}_{\mathrm {p,ggg}}(\triangle)&=\sum_{i,j,k} \wcen(M_i)\wcen(M_j)\wcen(M_k)w^{(3)}_{\mathrm {p,hhh}}(\triangle,M_i,M_j,M_k)\\
                    &+3\sum_{i,j,k} \wcen(M_i)\wsat(M_j)\wsat(M_k)w^{(3)}_{\mathrm{p,hpp}}(\triangle,M_i,M_j,M_k)\\
                    &+3\sum_{i,j,k} \wsat(M_i)\wcen(M_j)\wcen(M_k)w^{(3)}_{\mathrm{p,phh}}(\triangle,M_i,M_j,M_k)\\
                    &+\sum_{i,j,k} \wsat(M_i)\wsat(M_j)\wsat(M_k)w^{(3)}_{\mathrm{p,ppp}}(\triangle,M_i,M_j,M_k)
    \end{aligned}
\end{equation}

$r_{\mathrm{p12}},r_{\mathrm{p23}},r_{\mathrm{p31}}$ are abbreviated as $\triangle$. Solid lines on top panels of Fig.~\ref{fig:wp3} show the galaxy projected 3PCF computed using tabulation method. Similarly to the 2PCF the measurement from a single realization is noisier but consistent within expected errors for all tracers.

\subsection{Mass binning effects}
\label{sec:massbineffect}

We bin the mass of the host halo in 20 bins between around $11 < \log M_\star < 14.8$ (see App.\ref{app:downsamp} for detail about downsampling). The bins are narrow enough so that the hosting probabilities do not change significantly within the bins and assigning to each halo and particle a probability at the middle of the bin (we referred to this practice as a raw probability) is in most cases a good approximation. However, there are some exceptions if we shift satellite parameter $\log(M_0)$ of ELG just a little bit to $11.7$, where hosting probability drops very steeply below $\log(M_\mathrm{halo}) \sim 11.7$, the middle of bin failed to catch satellite information. Fig.~\ref{fig:tab} demonstrates the nature of this problem.

The black dash and dot-dash line on the top panel of Fig.~\ref{fig:tab} show the expected number of the central and satellite ELGs in our fiducial HOD model. Bold gray vertical dash line and gray vertical dot-dash lines show the edges and the middle of the mass bin. The peach and lavender histogram shows the number of halos and particles in each mass bin. Empty triangles pointing to the right show the raw weights based on the value in the middle of the mass bin and the filled triangle pointing to the left shows the refined weights. For most of the mass range, the two are very consistent. The last nonzero bin on the left ($4$th bin) is the exception. The mean number drops so steeply with the mass that it reaches an extremely low number for the middle mass of that bin. If we applied weight based on that value very few of the halos in that mass bin would acquire a satellite. This would incorrectly down-weight the halos close to the right edge of the mass bin that has a substantial probability of hosting a satellite. 

Increasing the number of mass bins is however impractical as it would significantly increase the number of separate cross-correlation functions that we need to keep track of. However, this problem would go away if we used a finer binning for the mass of the host halo. We modify our probabilities as

\begin{align}
    w^c_\mathrm{ref}(M_k) &= \displaystyle\sum_{k^{\prime}}^{N_{\mathrm{sub}}}\frac{N^{k,k^{\prime}}_h}{N^k_h} w_{\mathrm{raw}}^c(M_{k,k^{\prime}})
    =\displaystyle\sum_{k^{\prime}}^{N_{\mathrm{sub}}}\frac{N^{k,k^{\prime}}_h}{N^k_h}\langle N^{c}\rangle(M_{k,k^{\prime}}) \\
    w^s_\mathrm{ref}(M_k) &= \displaystyle\sum_{k^{\prime}}^{N_{\mathrm{sub}}}\frac{N^{k,k^{\prime}}_h}{N^k_h} w_{\mathrm{raw}}^s(M_{k,k^{\prime}})
    =\displaystyle\sum_{k^{\prime}}^{N_{\mathrm{sub}}}\frac{N^{k,k^{\prime}}_h}{N_p^k}\langle N^{s}\rangle(M_{k,k^{\prime}}) 
\end{align}

As shown in lower left panel on Fig.~\ref{fig:tab}, we further subdivided the mass bin into $N_{\mathrm{sub}}=20$ sub-bins, and take a weighted average of raw weights for each sub mass bin $w_{\mathrm{raw}}(M_{k,k^\prime})$ based on the number of halos $N^{k,k^{\prime}}_h$ in this sub mass bin. The dashed line on this panel shows the raw weight of this mass bin before correction, the solid line shows the refined weight. For central weight, the difference is subtle. For satellite weight, refined weight accurately accounts the contribution of satellites from this mass bin while raw weight failed to capture a number.

\begin{comment}
\begin{subequations} \label{eqn:wref}
\begin{align}
    &\wcen^{\mathrm{raw}}(M_{i,i^\prime})= p_{\mathrm{cen}}(M_{i,i^\prime})\\
    &\wsat^{\mathrm{raw}}(M_{i,i^{\prime}})= p_{\mathrm{sat}}(M_{i,i^\prime})\frac{N_{\rm halo}(M_i)}{N_{\rm part}(M_i)}\\
    &\wcen^{\mathrm{ref}}(M_{i})= \sum_{i^\prime}\wcen^{\mathrm{raw}}(M_{i,i^\prime})\frac{N_{\rm halo}(M_{i,i^\prime})}{N_{\rm halo}(M_i)}\\
    &\wsat^{\mathrm{ref}}(M_{i})= \sum_{i^\prime}\wsat^{\mathrm{raw}}(M_{i,i^\prime})\frac{N_{\rm halo}(M_{i,i^\prime})}{N_{\rm halo}(M_i)}
\end{align}
\end{subequations}
\end{comment}
The lower right panel of Fig.~\ref{fig:tab} compares the tabulated projected 2PCF computed with the raw weights and the refined weights. The raw weights clearly fail to describe the 2PCF on the scales of $r_p < 0.1$ where the systematic offsets are more than $50 \%$ of the signal.

This problem will appear whenever there are variations in the expected number of galaxies (either satellites or centrals) across the width of the bin. When assigning all galaxies in the bin the weight at the middle of the bin we are making an approximation
\begin{equation}
    \displaystyle\int\displaylimits_{M_{\mathrm{min}}}^{M_{\mathrm{max}}}\!\!\!\mathrm{d}M\,w(M)\frac{dN}{dM} \approx w(\overline{M})N^\mathrm{tot},
\end{equation}
where $\overline{M}$ is the middle of the bin, $\Delta M$ is the width of the bin, the integration is between $\overline{M}-\Delta M/2$ and $\overline{M}+\Delta M/2$, and the $N^\mathrm{tot}$ is the total number of galaxies in that bin. To see when this approximation may fail we can expand the weight function on the left side of the equation in Taylor series around the middle point as
\begin{align}
    w(M) &= w(\overline{M}) + \frac{dw}{dM}\Big|_{M=\overline{M}}(M-\overline{M}) + \frac{1}{2}\frac{d^2w}{dM^2}\Big|_{M=\overline{M}}(M-\overline{M})^2 + \nonumber \\
    & \mathcal{O}[(M - \overline{M})^3].
\end{align}
The integral over the first term results in $w(\overline{M})N^\mathrm{tot}$, which matches exactly with our approximation. The second term integrates to zero (as an odd function over symmetric limits). The third term integrates to $w''(\overline{M})(\Delta M)^3 N^\mathrm{tot}/24$, here the double prime denotes the second derivative with respect to halo mass, $\Delta M$ is the width of the bin.\footnote{We are assuming that the fundamental quantities such as the number of halos and the bias of halos does not change significantly within the bin. If this is not the case, the binning is obviously too broad and needs to be refined.} This is the leading error term and it needs to be small for the approximation to be valid. The condition is

\begin{equation}
    \frac{w''(\overline{M})}{w(\overline{M})} \ll \frac{24}{(\Delta M)^3}
\end{equation}

If this condition between the second derivative of the weights and the width of the mass bin starts breaking down the weight refining procedure described in this section may be in order. This condition is obviously broken for our HOD models since they contain a discontinuous function in the satellite probability in eqs.~\eqref{eq:Nsatlrg} and \eqref{PsateELG}).

\setlength{\tabcolsep}{2pt}
\begin{table*}
\begin{tabular}{|p{30pt}*{9}{|>{\centering\arraybackslash\hspace{0pt}}p{48pt}}|}
\cline{1-10}
Tracer & \multicolumn{3}{c|}{LRG} &
\multicolumn{3}{c|}{ELG} &
\multicolumn{3}{c|}{QSO}\\
\cline{1-10}
Data &$\wptwo$ &$w_{\mathrm{p(SV)}}^{(3)}$ &$\wptwo+w_{\mathrm{p(SV)}}^{(3)}$&$\wptwo$ &$w_{\mathrm{p(SV)}}^{(3)}$ &$\wptwo+w_{\mathrm{p(SV)}}^{(3)}$&$\wptwo$ &$w_{\mathrm{p(SV)}}^{(3)}$ &$\wptwo+w_{\mathrm{p(SV)}}^{(3)}$\tabularnewline
\cline{1-10}
$\log{M_{\mathrm{cut}}}$&$12.88\pm0.199$&$12.73\pm0.058$&$12.73\pm0.059$
                        &$11.83\pm0.059$&$11.74\pm0.125$&$11.83\pm0.054$
                        &$12.47\pm0.060$&$12.43\pm0.130$&$12.48\pm0.058$
\tabularnewline
$\sigma$                &$0.315\pm0.200$&$0.151\pm0.103$&$0.162\pm0.105$
                        &-              &-              &-
                        &-              &-              &-
\tabularnewline
$\log{M_1}$             &$13.93\pm0.141$&$13.83\pm0.053$&$13.82\pm0.047$
                        &-              &-              &-
                        &$15.49\pm0.766$&$15.53\pm0.851$&$15.53\pm0.755$
\tabularnewline
$\log{M_0}$             &$11.73\pm0.431$&$11.76\pm0.407$&$11.78\pm0.400$
                        &-              &-              &-
                        &-              &-              &-
\tabularnewline
$\alpha$                &$1.279\pm0.055$&$1.300\pm0.040$&$1.301\pm0.038$
                        &$0.188\pm0.097$&$0.268\pm0.161$&$0.178\pm0.095$
                        &-              &-              &-
\tabularnewline
$A_{\mathrm{c}}$        &-              &-              &-              
                        &-              &-              &-
                        &-              &-              &-
\tabularnewline
$A_{\mathrm{s}}$        &-              &-              &-
                        &$0.015\pm0.007$&$0.023\pm0.016$&$0.015\pm0.006$
                        &-              &-              &-
\tabularnewline
$\chi^2/\mathrm{d.o.f}$ &$2.5/(12-5)$   &$36.5/(99-5)$  &$34.9/(111-5)$
                        &$0.9/(11-3)$   &$57.5/(85-3)$  &$56.7/(96-3)$
                        &$4.3/(9-2)$    &$39.4/(60-2)$  &$42.6/(69-2)$
\tabularnewline
\cline{1-10}

\end{tabular}
\caption{The results for the fits to HOD mock catalog of each tracers with three different data set: 2PCF only ($\wptwo$), 3PCF only ($w_{\mathrm{p(SV)}}^{(3)}$) and joint ($\wptwo+w_{\mathrm{p(SV)}}^{(3)}$). We shows the $\mathrm{mean}\pm 1\sigma$ error for floating HOD parameters and $\chi^2/\mathrm{d.o.f}$ for each fits. The fiducial (true) value for each parameter are listed in Tab. \ref{tab:fidparams}}
\label{tab:margstat}
\end{table*}

\begin{figure}
    \centering
    \includegraphics[width=0.9\linewidth]{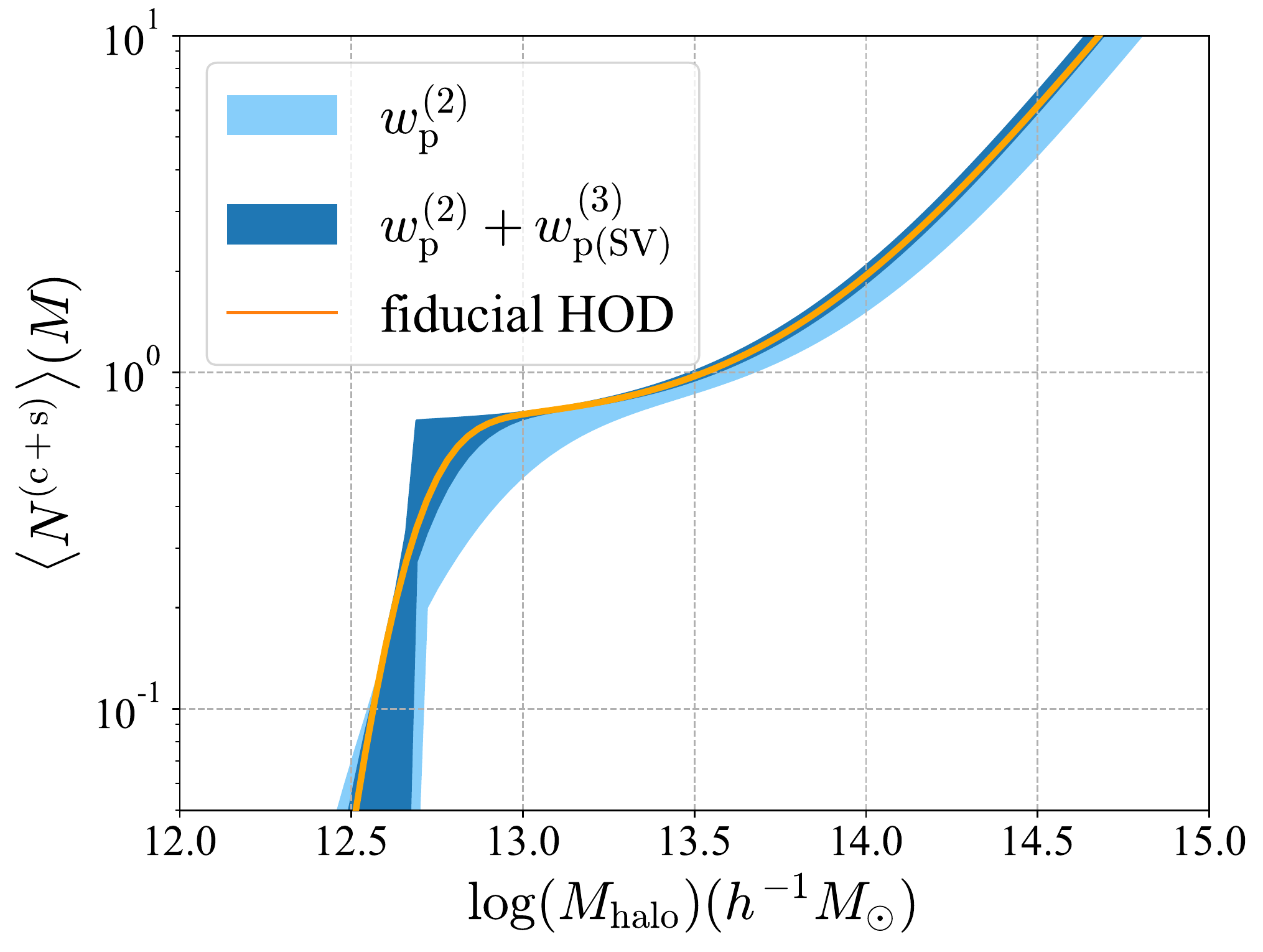}
    \caption{\protect\footnotesize 1-$\sigma$ band of LRG sample HOD. The light blue is the 68$\%$ CL uncertainty from projected 2PCF only, the dark blue band is the 68$\%$ CL uncertainty from joint fitting of projected 2PCF and simplified version projected 3PCF. Orange line is the fiducial HOD of LRG sample.}
    \label{fig:hodbandlrg}
\end{figure}

\begin{figure*}
\centering
\includegraphics[width=0.9\linewidth]{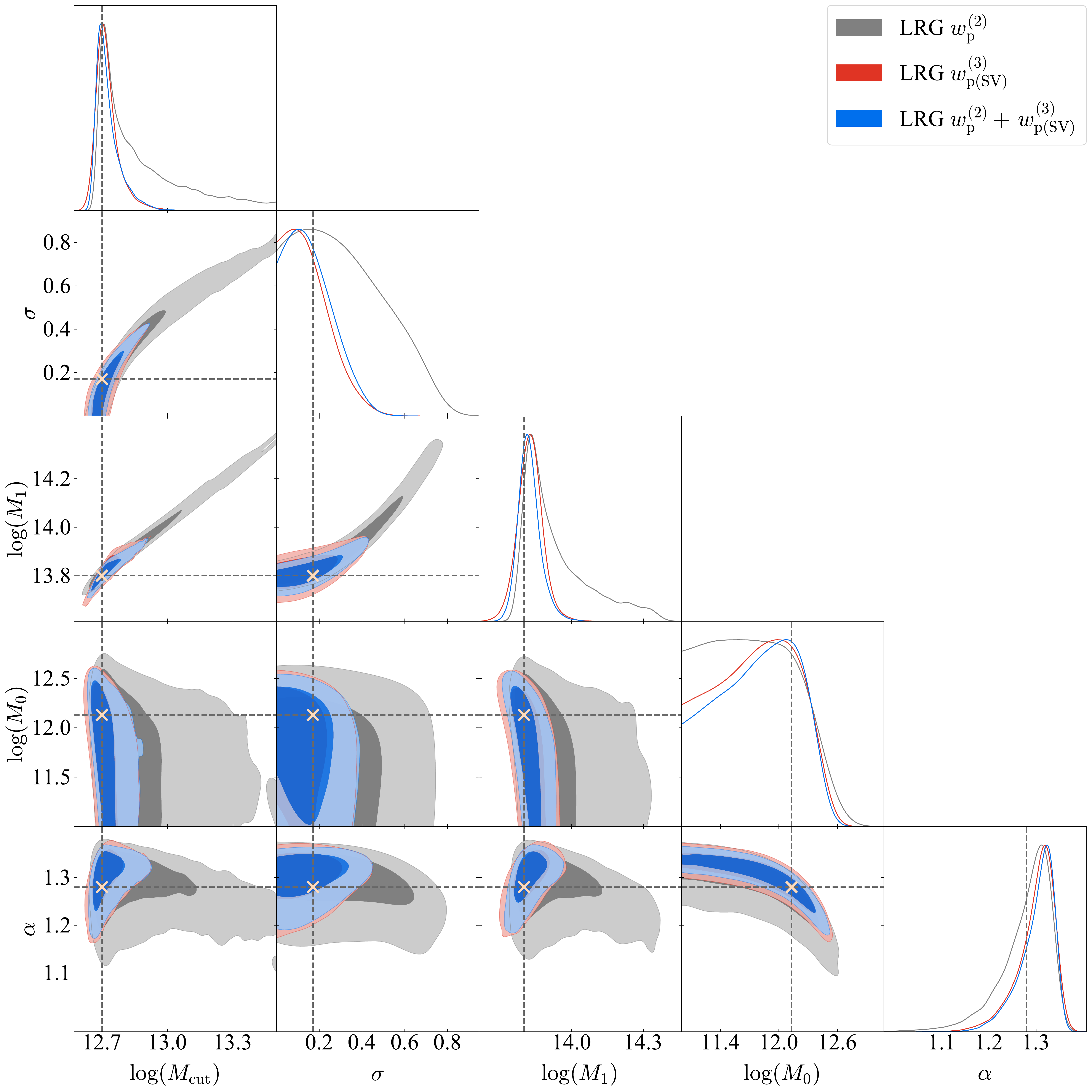}
\caption{\protect\footnotesize Marginalized probability distribution of HOD parameters for DESI like LRG sample at $z=0.8$. The results from the projected 2PCF and 3PCF are shown in grey and red respectively. Blue shows the joint constraints from the two. The contours represent 68 and 95 percent confidence levels. 1D marginalized distribution for each parameters are shown on top of each column. The dash line and light yellow cross markers shows fiducial HOD parameter values.}
\label{fig:LRGcontour}
\end{figure*}

\begin{figure*}
\centering
\includegraphics[width=0.9\linewidth]{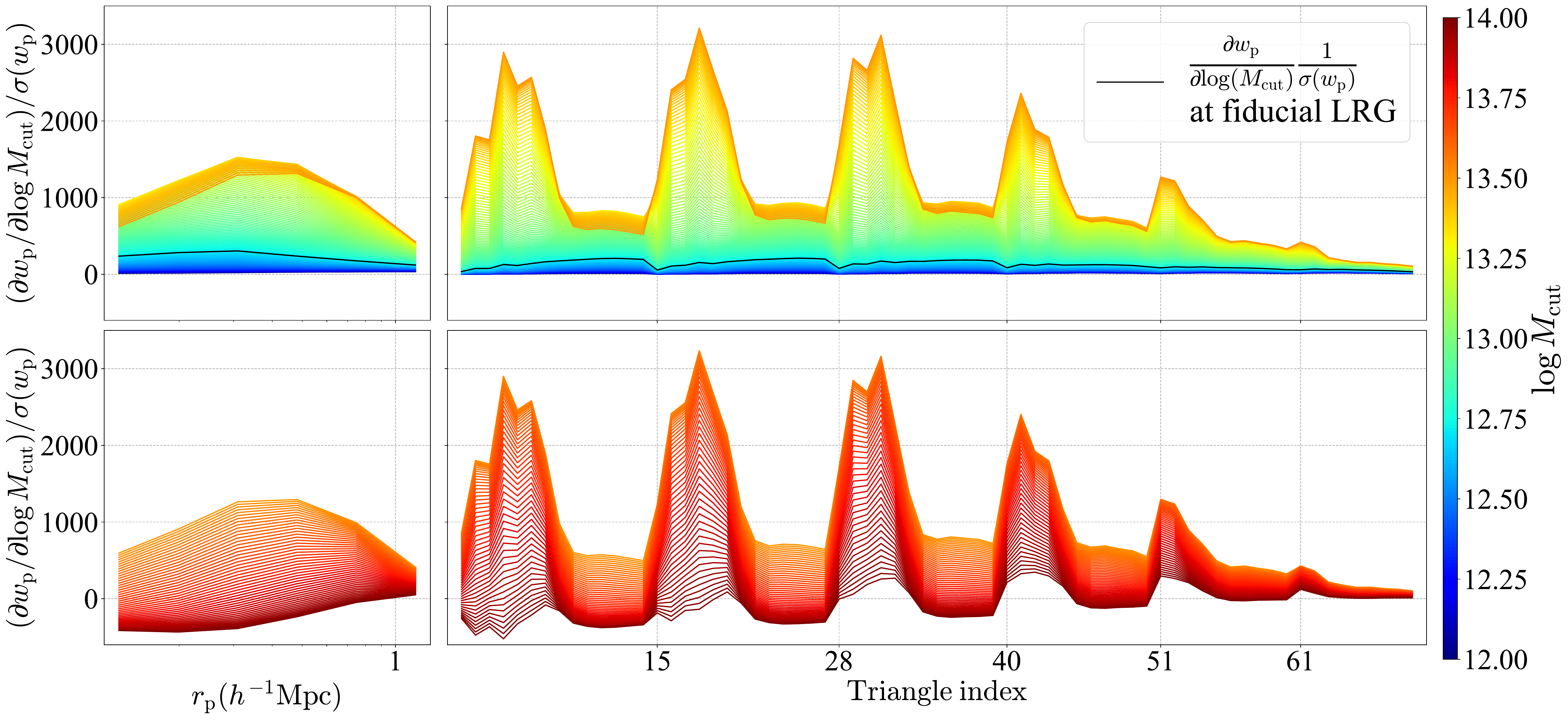}
\caption{\protect\footnotesize Partial derivative of $w_{\mathrm{p}}$ with respect to $\log(M_{\mathrm{cut}})$ normalized using error of $w_{\mathrm{p}}$ when fixing other HOD parameters. Plot has been separated to two panels to avoid overlap when partial derivative drop down.}
\label{fig:LRGderivative}
\end{figure*}

\begin{figure*}
\centering
\includegraphics[width=0.9\linewidth]{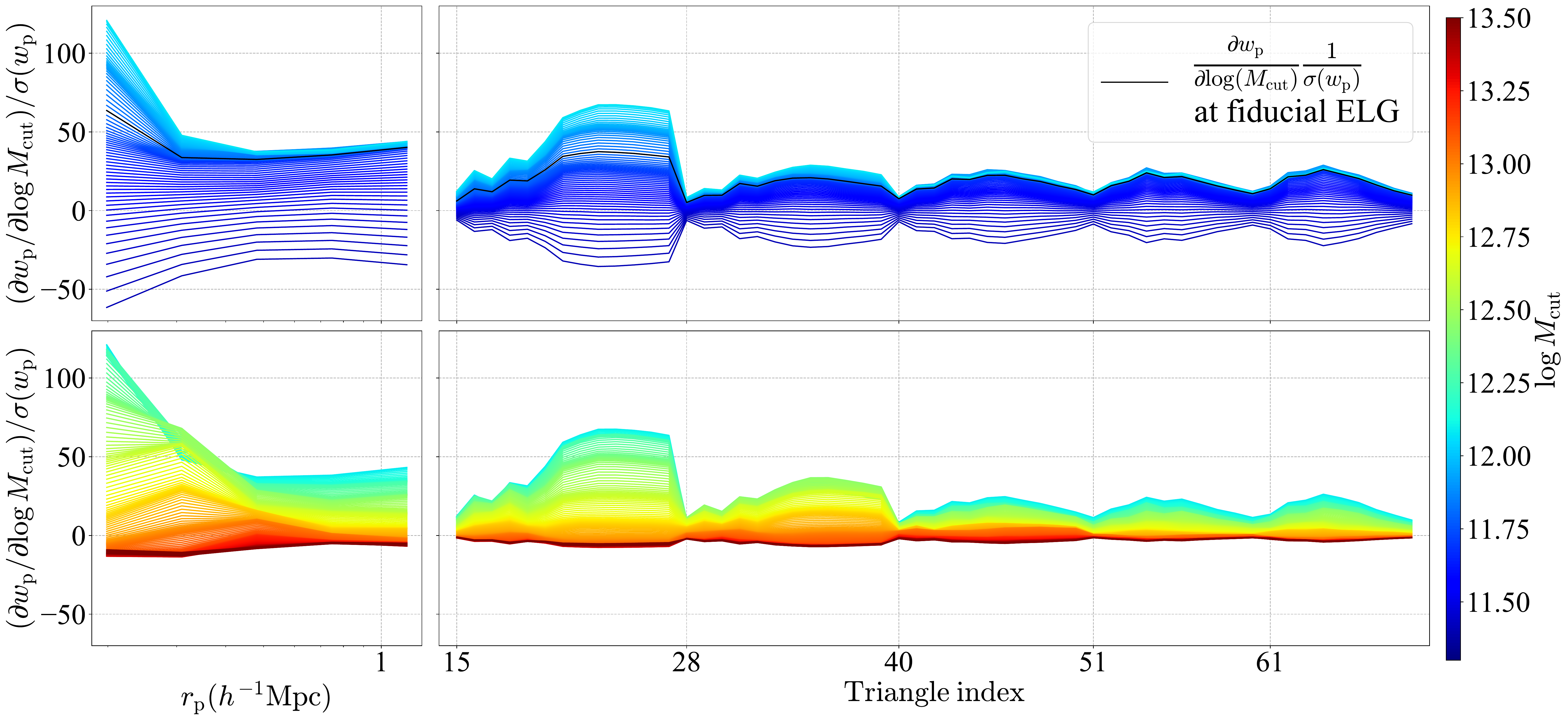}
\caption{\protect\footnotesize Similar plot as Fig.~\ref{fig:LRGderivative} for ELG sample at fiducial redshift $z=1.1$}
\label{fig:ELGderivative}
\end{figure*}

\begin{figure}
\centering
\subfigure {\includegraphics[width=\columnwidth]{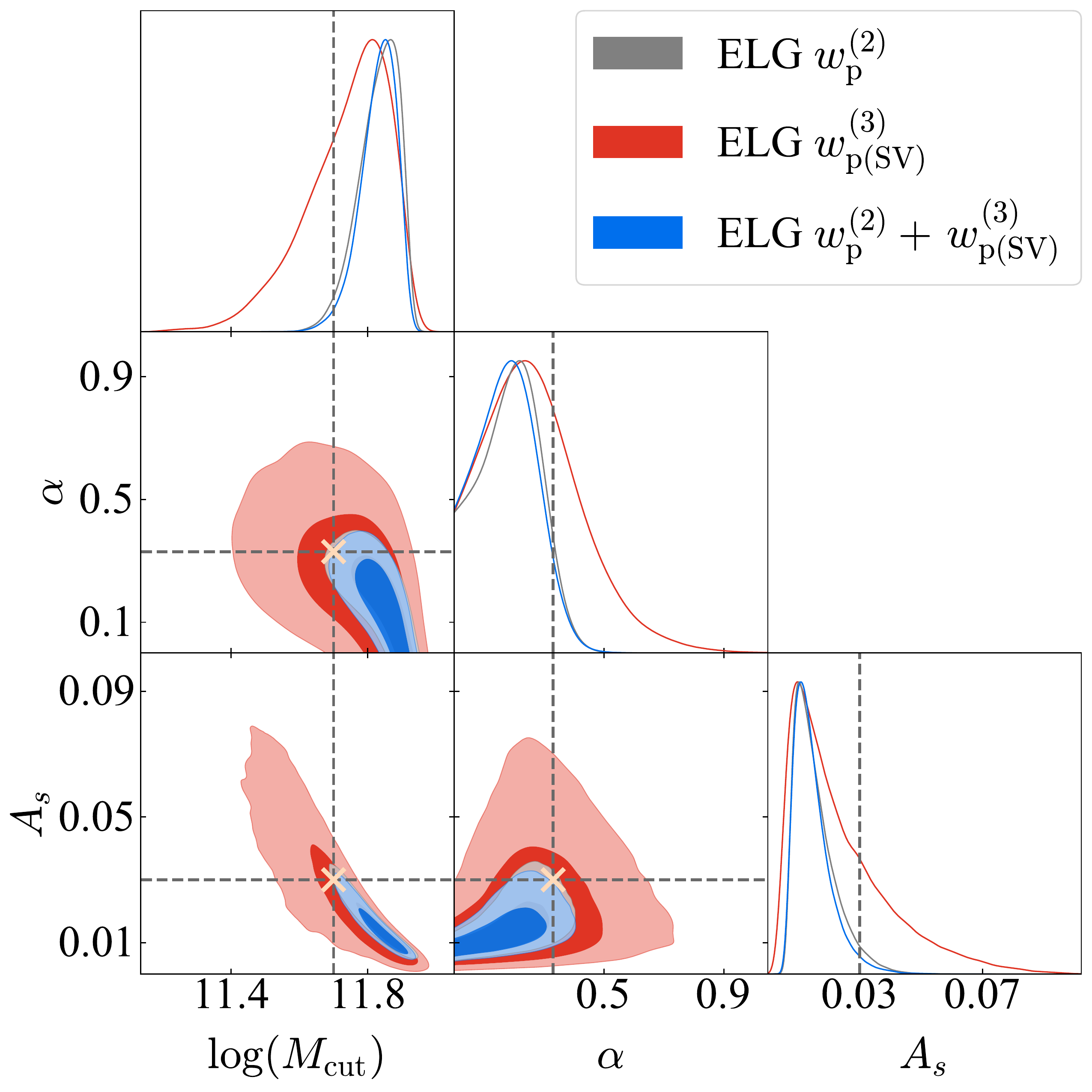}}
\subfigure {\includegraphics[width=\columnwidth]{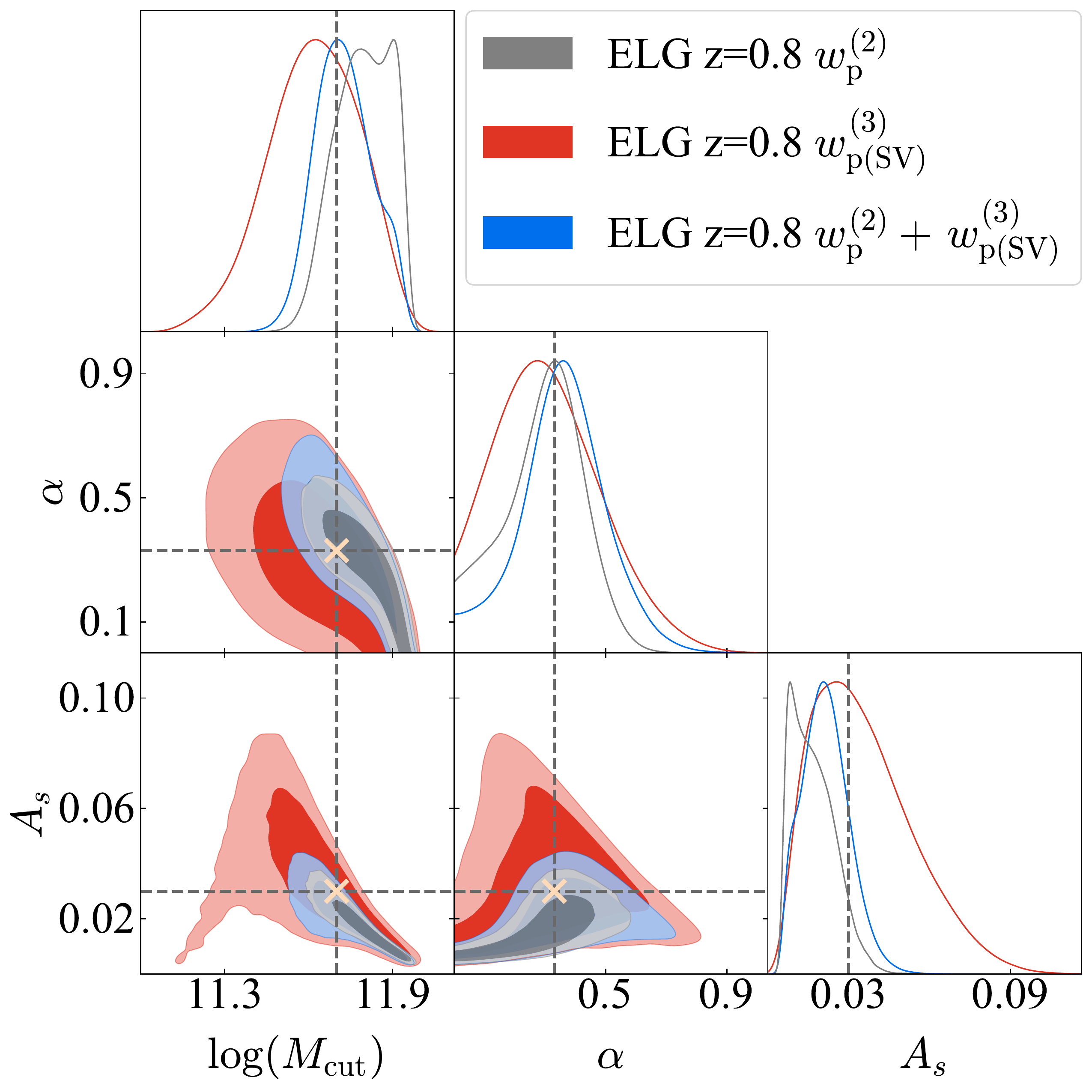}}
\caption{\protect\footnotesize Marginalized probability distribution of selected HOD parameters for DESI like ELG sample at $z=1.1, 0.8$. The results from the projected 2PCF and 3PCF are shown in grey and red respectively. Blue shows the joint constraints between the two. The contours represent 68 and 95 percent confidence levels. 1D marginalized distribution for each parameter is shown on top of each column. The dashed line and light yellow cross markers shows fiducial HOD parameter values.}
\label{fig:ELGcontour}
\end{figure}

\begin{figure}
\centering
\includegraphics[width=0.9\linewidth]{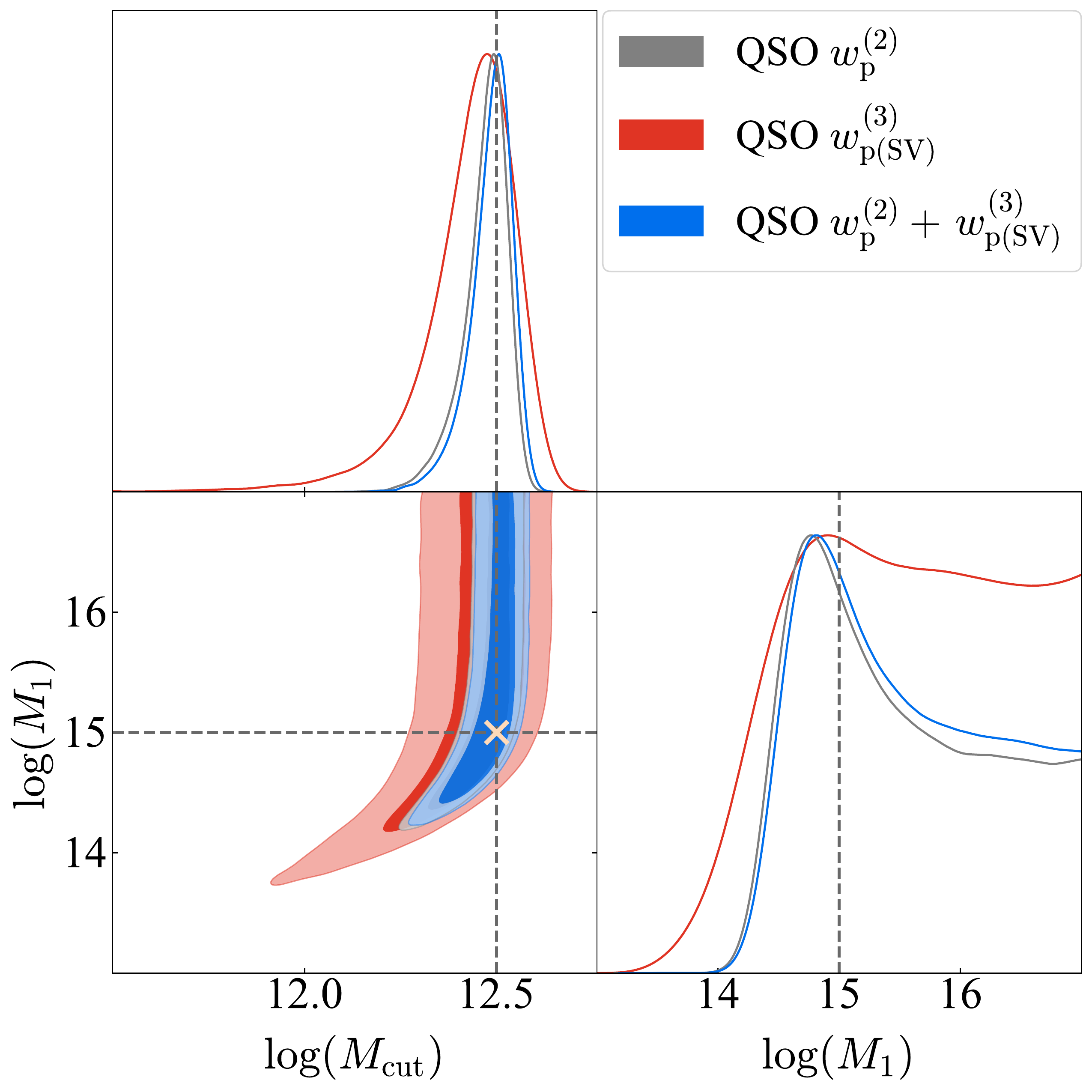}
\caption{\protect\footnotesize Marginalized probability distribution of selected HOD parameters for DESI like QSO sample at $z=1.4$. The results from the projected 2PCF and 3PCF are shown in grey and red respectively. Blue shows the joint constraints between the two. The contours represent 68 and 95 percent confidence levels. 1D marginalized distribution for each parameter is shown on top of each column. The dashed line and light yellow cross markers shows fiducial HOD parameter values.}
\label{fig:QSOcontour}
\end{figure}

\section{Results}

We create DESI like LRG, ELG, and QSO samples as described in Sec.~\ref{sec:mock}. We then use the MCMC method to fit the HOD parameters for the model described in Sec.~\ref{sec:mcmc} with the covariance matrix obtained as described in Sec.~\ref{sec:covmat}. The covariance matrix represents the variance in the measurements expected from a cosmic volume of 1 cubic GigaParsecs. The actual DESI measurements will be obtained from larger volumes, but since the errors on both the 2PCF and the 3PCF scale similarly with the volume the relative strength of the constraints coming from the two will not change.

Fig.~\ref{fig:hodbandlrg} shows 1 $\sigma$ uncertainty band of LRG sample HOD function from 2PCF only fitting and 2PCF+3PCF joint fitting. The light blue band shows $68\%$ CL uncertainty from 2PCF only and the dark blue band shows the band from 2PCF+3PCF joint fitting. The orange line represents the fiducial HOD setting as the truth behind the mock we fit to. It is clear to see joint fitting has a much narrow band compared to the one using 2PCF only, especially for the range $\log(M_{\mathrm{halo}} > 12.7)$, indicate a much better constraint on satellite parameters from joint fitting. Fiducial HOD lie in the $1\sigma$ band shows a good recovery for both cases.

Fig.~\ref{fig:LRGcontour} shows 1 and 2 $\sigma$ confidence level contours on the HOD parameters for the LRG sample. These constraints are dominated by the $w_{\mathrm{p(SV)}}^{(3)}$. The improvement is especially large for the parameters $\log M_\mathrm{cut}$, $\sigma$, and $\log M_1$. The 3PCF constraints on those parameters improve by 70, 49, and 62 percent respectively compared to the 2PCF results. Combined fitting does not significantly differ from the 3PCF only results. Tab.~\ref{tab:margstat} summarizes the marginalized statistic for each fit. From the 1D distribution of each parameter on Fig.~\ref{fig:LRGcontour}, all cases successfully recover the fiducial HOD parameters. 

Fig.~\ref{fig:ELGcontour} and \ref{fig:QSOcontour} show 1 and 2 $\sigma$ confidence level contours for the ELG at redshift $1.1$ and $0.8$ and QSO samples at redshift $1.4$ respectively. We only free HOD parameters as shown in the contours for these tracers. For the ELG and QSO, the constraints are dominated by the projected 2PCF. Improvements offered by the addition of the projected 3PCF are negligible. 

There could be several reasons why the LRGs benefit greatly from the addition of the 3PCF information while ELGs and QSOs do not. One potential explanation is that the ELGs and QSOs are at higher redshifts where matter underwent less nonlinear evolution and the three-point signal is not as pronounced. Another potential explanation is that galaxies of different host halo masses are not equally sensitive to the three-point information  \citep[see e.g.][]{2007MNRAS.378.1196K}.

To study the sensitivity of 2PCF and 3PCF to HOD parameters at different fiducial values we make a plot of the partial derivative of $\wptwo$ and $w_{\mathrm{p(SV)}}^{(3)}$ with respect to $\log(M_\mathrm{cut})$ normalized to the variance in the measurement at the fiducial value.  Fig.~\ref{fig:LRGderivative} shows partial derivatives of the 2PCF and the 3PCF with respect to $\log(M_\mathrm{cut})$ with other parameters fixed to their fiducial value. To make the plot more readable we separate it into two parts. The top panel covers the range $12<\log(M_\mathrm{cut})<13.5$ while the bottom panel covers $13.5<\log(M_\mathrm{cut})<14$. High values of this derivative mean that the measurement at that specific bin is highly sensitive to small changes in $M_\mathrm{cut}$

The derivative of $\wptwo$ reaches highest value at $\log(M_\mathrm{cut})=13.28$ then drops back, while derivative of $w_{\mathrm{p(SV)}}^{(3)}$ keeps increasing up until $13.5$ and only then drops down. The 3PCF displays a larger cumulative sensitivity in the range $\log(M_\mathrm{cut})>13.16$, below that range the 3PCF is not as sensitive to small changes in $M_{\mathrm{cut}}$ compare to $\wptwo$.

Another thing apparent from the figure is that the small scale triangles are more sensitive to $\log(M_\mathrm{cut})$ compared to their large-scale counterparts (as evident by the local peaks in the right panel). Triangles with all side lengths within the first 6 bins (side lengths $r_{\mathrm{p}}<1.41\hMpc$, corresponding to triangle indices 1-8, 15-21, 28-33, 40-43, 51-53, 61), peak at $\log(M_\mathrm{cut})=13.5$, while other triangles behave just like $\wptwo$, drop back at $\log(M_\mathrm{cut})=13.28$. The different behavior of small scale triangles and small scale pairs leads to a higher sensitivity to HOD parameter changes for small scale $w_{\mathrm{pSV}}^{(3)}$. The normalized derivative of $w_{\mathrm{p(SV)}}^{(3)}$ hit around 3000 while $\wptwo$ remains at 1500.

Fig.~\ref{fig:ELGderivative} shows the similar plots for ELG sample at $z=1.1$. The sensitivity in both the 2PCF and the 3PCF increases in the range of $11.3<\log(M_\mathrm{cut})<11.98$ and then drops in the range of $11.98<\log(M_\mathrm{cut})<13.5$. At this redshift, the top sensitivity is achieved at the values of around $\log(M_\mathrm{cut})=11.98$. The sensitivity of $\wptwo$ at the top is higher than the sensitivity of the $w_{\mathrm{p(SV)}}^{(3)}$. For ELGs, means a lower sensitivity for 3PCF. The cumulative sensitivity at the peak is also larger for the 2PCF compared to the 3PCF. Small scale triplets do not show the same behavior as the LRG sample, remaining at low sensitivity compare to small scale pairs.

These two plots show that both the redshift and the typical halo mass are responsible for the difference between the LRG and the ELG cases. The DESI LRGs happen to be in the halo mass range where the 3PCF is more sensitive to the HOD parameters, while ELGs are in the halos with the opposite property. This is the main reason why the improvement in our ELG constraints is modest while the improvement in the LRG constraints is significant.

Plots similar to the ones presented in Figs.~\ref{fig:LRGderivative} and \ref{fig:ELGderivative} can be used to determine whether the 3PCF is expected to affect the overall HOD constraints in a meaningful way. For example, one could populate the simulation with the best fit HOD parameters resulting from the 2PCF fits and do the same thing with slightly offset values of HOD parameters. Derivatives of 3PCF with respect to HOD parameters can be estimated by computing the 3PCF for these mocks (using finite differences). The covariance matrix (standard error) can be computed using the jackknife method Unlike MCMC chains, that typically require order of 10,000 computations, this can be done with just a handful 3PCF computations (a few for the numerical derivatives and order of 100 3PCF computations for the jackknife covariance) and can be done without preparing the tabulated triplet counts. A comparison of these integrated sensitivities of the 3PCF and the 2PCF can then be used to decide whether the additional investment of computational resources for the computation of the tabulated triplet counts is warranted.

To test the pure redshift dependence of the constraining power of the 3PCF we populate our ELG mock catalog at redshift $0.8$ with the same HOD parameters (tuned to the same number density, i.e. ratio of $A_c$ and $A_s$ remain unchanged). The results are presented on the bottom panel of Fig.~\ref{fig:ELGcontour}. We do not find a significant change in the overall picture. The constraints are still dominated by the 2PCF signal. We do notice however that the addition of the 3PCF makes the likelihood contours more Gaussian and moves the most likely values closer to the true values somewhat debiasing the results.

\section{Conclusion}

We studied the performance of projected 3PCFs in constraining HOD parameters for different mock galaxy samples , which based on \textsc{AbacusSummit} simulation, targeted by DESI. We generalized the tabulation method to 3PCF computations to make a fast evaluation of the posterior likelihood possible. 

We find that the constraints on the basic HOD parameters of the mock LRG sample with input HOD parameter as shown in Tab.\ref{tab:fidparams} at redshift $z \sim 0.8$ can be significantly improved by the addition of the 3PCF. The constraints on some parameters have improved by as much as 70 $\%$. For the characteristic minimum mass of the central LRGs we get the constraints $\log(M_\mathrm{cut})=12.88\pm0.199$ with the 2PCF and $\log(M_\mathrm{cut})=12.73\pm0.058$ with the 3PCF. For the threshold mass of the satellite LRGs we get the constraints $\log(M_1)=13.93\pm0.141$ with the 2PCF and $\log(M_1)=13.83\pm0.053$ with the 3PCF. All at 1$\sigma$ confidence level.

We also find that the additional constraining power offered by the 3PCF depends on the redshift of the galaxy sample as well as the typical halo mass that its galaxies occupy. The relative strength of the 3PCF increases at lower redshifts. 3PCF is also a more sensitive measurement for the samples that incorporate more massive halos. The ELG samples of DESI are at higher redshifts and occupy less massive halos. This results in the 3PCF not being as efficient in constraining their host halo mass ranges. For the mock ELG sample  with input HOD parameter as shown in Tab.\ref{tab:fidparams} at redshift $z \sim 1.1$ the constraints of the characteristic minimum mass of the central are $\log(M_\mathrm{cut})=11.83\pm0.059$ with the 2PCF and $\log(M_\mathrm{cut})=11.74\pm0.125$ with the 3PCF. For the mock QSOs with lower number density compare to the other tracers and even higher redshift $z \sim 1.4$, we get $\log(M_\mathrm{cut})=12.47\pm0.060$ with the 2PCF and $\log(M_\mathrm{cut})=12.43\pm0.130$ with the 3PCF, 2PCF remaining dominates. 

\section*{Acknowledgements}
We would like to thank Gongbo Zhao, Shun Satio, Hee Jong Seo, Andrew Hearin, Francisco Villaescusa-Navarro, Ashley J. Ross, and Lehman Garrison for their helpful discussion.
LS is grateful for support from DOE grants DE-SC0021165
and DE-SC0011840, NASA ROSES grants 12-EUCLID12-0004
and 15-WFIRST15-0008, and Shota Rustaveli National Science
Foundation of Georgia grants FR 19-498.

This research is supported by the Director, Office of Science, Office of High Energy Physics of the U.S. Department of Energy under Contract No. DE–AC02–05CH11231, and by the National Energy Research Scientific Computing Center, a DOE Office of Science User Facility under the same contract; additional support for DESI is provided by the U.S. National Science Foundation, Division of Astronomical Sciences under Contract No. AST-0950945 to the NSF’s National Optical-Infrared Astronomy Research Laboratory; the Science and Technologies Facilities Council of the United Kingdom; the Gordon and Betty Moore Foundation; the Heising-Simons Foundation; the French Alternative Energies and Atomic Energy Commission (CEA); the National Council of Science and Technology of Mexico; the Ministry of Economy of Spain, and by the DESI Member Institutions.

We acknowledge the use of the NASA astrophysics data system
https://ui.adsabs.harvard.edu/ and the arXiv open-access repository
https://arxiv.org/.
The software was hosted on the GitHub platform https://github.com/. 
The manuscript was typeset using the overleaf cloud-based LaTeX editor https://www.overleaf.com.

%%%%%%%%%%%%%%%%%%%%%%%%%%%%%%%%%%%%%%%%%%%%%%%%%%
\section*{Data Availability}

The data product related to this study, including tabulated 2 point and 3 point counts, HOD mock catalogs, jackknife covariance matrices, and MCMC chains, is available at \url{https://doi.org/10.5281/zenodo.6380446}.

The AbacusSummit simulations used in this study are publicly available (\url{https://abacusnbody.org/}).


%\bibliographystyle{apa} 
%\bibliographystyle{mnras}
%\bibliography{key,intro} % if your bibtex file is called example.bib
\input{hod.bbl}

\newpage

%%%%%%%%%%%%%%%%%%%%%%%%%%%%%%%%%%%%%%%%%%%%%%%%%%

%%%%%%%%%%%%%%%%% APPENDICES %%%%%%%%%%%%%%%%%%%%%

\appendix

\section{Downsampling}
\label{app:downsamp}

The number of halos in simulations increases rapidly towards the lower mass range. However, the contribution of low mass halos to the 2PCF and 3PCF is negligible for the hod parameter range we are interested in (see Fig.~\ref{fig:hodmodel}).

When populating mocks, we set a hard cut-off mass for halo samples at the lower mass end. We remove all halos with a mass less than $10^{11} h^{-1}M_{\odot}$. 

\begin{figure}
\centering
\includegraphics[width=0.9\linewidth]{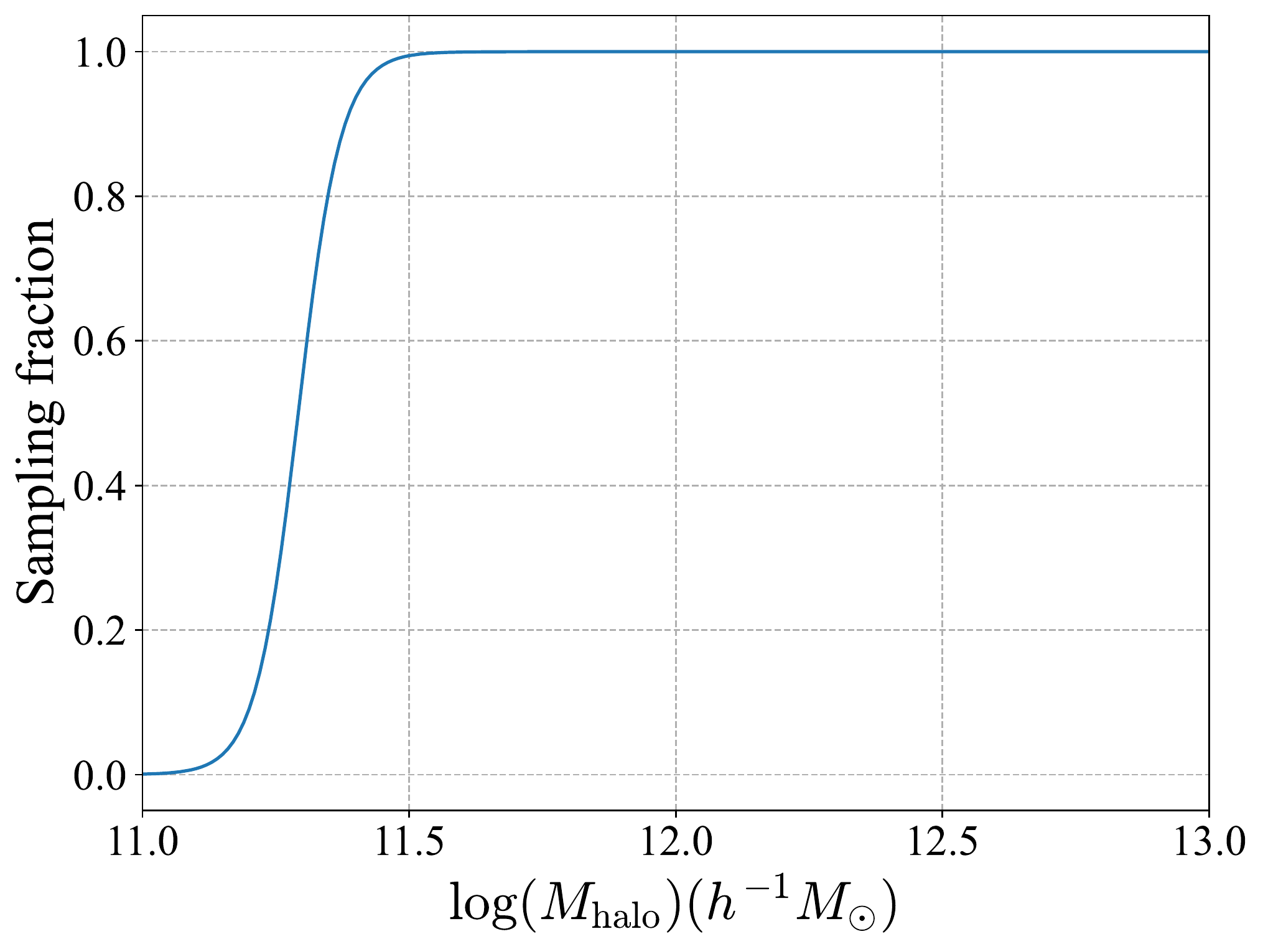}
\caption{\protect\footnotesize Fraction of halos we take from all halo as a function of halo mass.}
\label{fig:downsample}
\end{figure}

For the tabulation method, to speed up pair and triangle counting when preparing tables, we downsample the lower mass halos using the filter,
\begin{equation}
\label{eqn:down}
    \mathrm{frac_{halos}}(M_\mathrm{halo})=\frac{1}{1 + 10\exp{(-25\log(M_\mathrm{halo}) - 11.2))}}
\end{equation}
from the \textsc{AbacusHOD} package \citep{2021arXiv211011412Y}. 
Fig.~\ref{fig:downsample} shows the halo sample fraction as a function of halo mass following Eqn. \ref{eqn:down}. This filter will further remove excess halos that have trivial contribution to the clustering in low mass bins.

For the particles in table preparation process, we randomly select $0.15\%$ of particles making each halo for the pair and triangle counting. This small percentage is still larger than the mean satellite number for each halo. 

We explicitly checked that the clustering from the tabulation method, which applied the filter, could perfectly match the clustering from mocks, which applied only a hard cut-off, around the fiducial HOD parameter area we are interested in (see Fig.~\ref{fig:wp2} and \ref{fig:wp3}).

ELGs have a lower typical host halo mass compared to other samples and they could in principle be affected by the filter. We omit the first separation bin for the ELGs to protect against this eventuality for some extreme HOD parameter values.

\section{Choice of maximum radial separation}
\label{app:pistar}

\begin{figure*}
\centering
\includegraphics[width=0.9\linewidth]{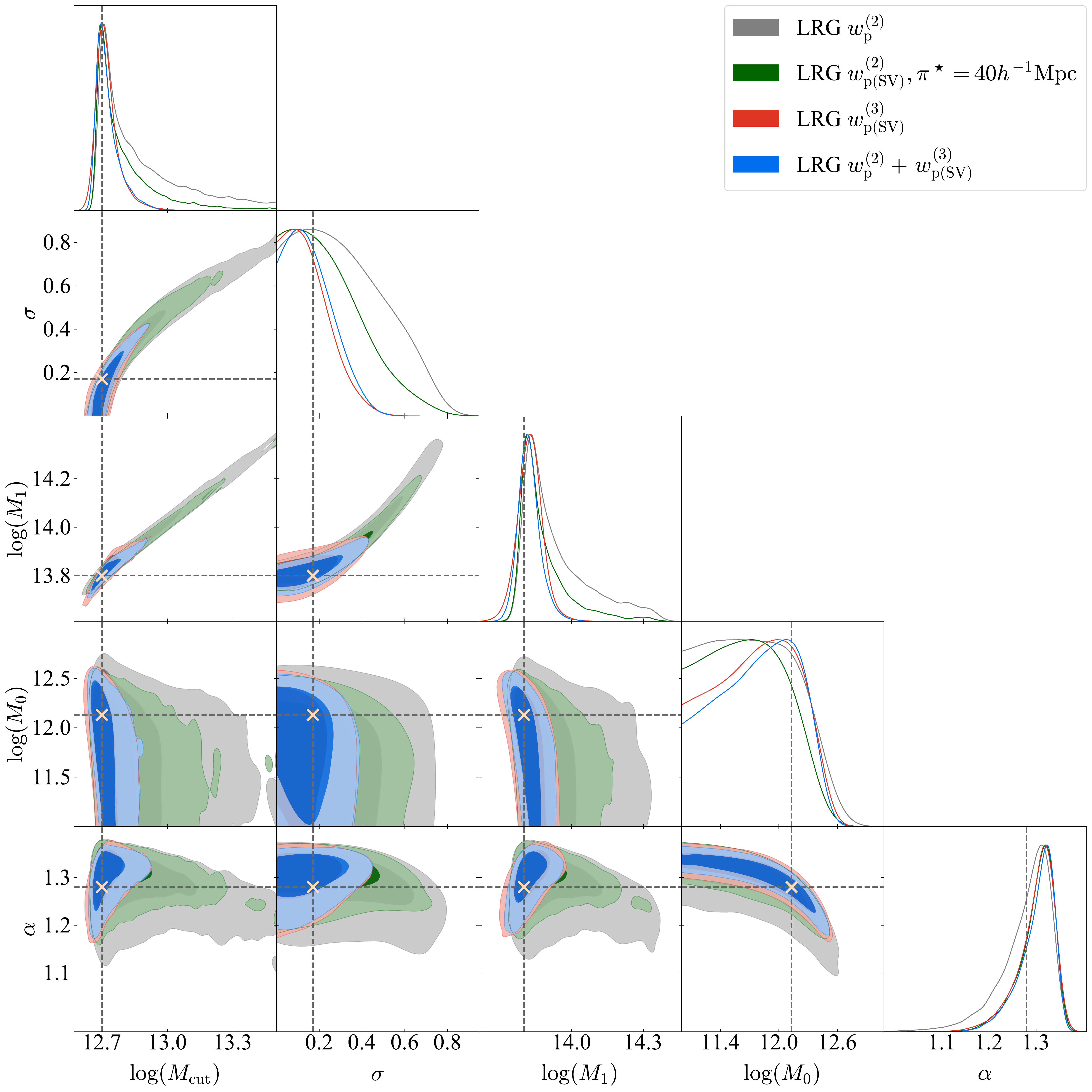}
\caption{\protect\footnotesize A similar plot as \ref{fig:LRGcontour} with additional green contours shows results from the projected 2PCF but with a value of $\pi^\star = 40h^{-1}$ Mpc. }
\label{fig:pistar}
\end{figure*}

The choice of $\pi^\star$ value affects the resulting constraints on HOD parameters. To minimize RSD effect and make the comparison more direct to simplified projected 3PCF, which ignored line-of-sight separation, we extend this value to 100$h^{-1}$ Mpc in the main analysis. To check that this does not significantly alter results we also ran MCMC chains where this value was set to a more conventional $\pi^\star = 40h^{-1}$ Mpc for the projected 2PCF. These results are presented in Fig.~\ref{fig:pistar}. The figure is identical to Fig.~\ref{fig:LRGcontour} except we added green contours that correspond to the projected 2PCF results with $\pi^\star = 40h^{-1}$ Mpc. This choice of $\pi^\star$ is indeed more optimal but the likelihood surfaces do not change enough to alter any of our main conclusions. The projected 3PCF(SV) still dominates the joint constraints. Applying full definition projected 3PCF would likely also increase its constraining power.

\section{Analytical Random Triplet Counts}
\label{app:anarand}

\begin{figure}
\centering
\includegraphics[width=0.9\linewidth]{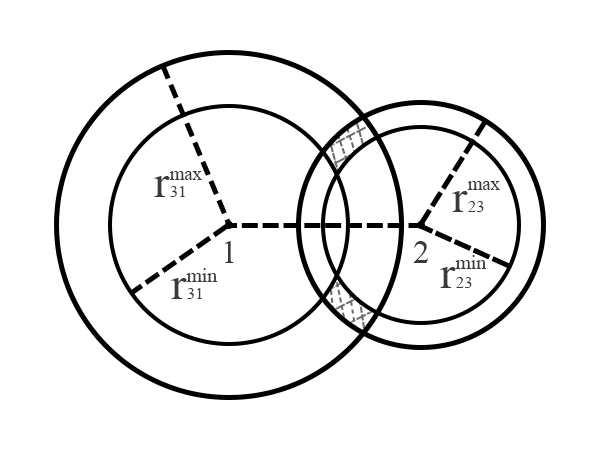}
\caption{\protect\footnotesize Geometry of random triplets problem. A fixed pair $r_{12}$ with certain binning setting can only have triplets shown in the shaded area.}
\label{fig:RRRanalytic}
\end{figure}

\begin{figure}
\centering
\includegraphics[width=0.9\linewidth]{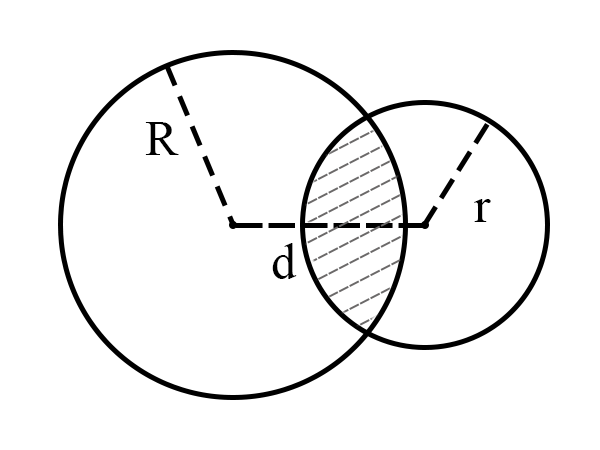}
\caption{\protect\footnotesize Intersection of two circles with radius $R$, $r$ and distance of centers $d$. Intersection area A has been shaded.}
\label{fig:circleintersect}
\end{figure}

The $RRR\left(r_{12}^\mathrm{min},r_{12}^\mathrm{max},r_{23}^\mathrm{min},r_{23}^\mathrm{max},r_{31}^\mathrm{min},r_{31}^\mathrm{max}\right)$ represent the average number of triplets of a random (spatially uncorrelated) distribution of galaxies, where the perpendicular to the line-of-sight distance between the triplet points satisfies conditions $r_{12}^\mathrm{min} < r_{p12} < r_{12}^\mathrm{max}$, $r_{23}^\mathrm{min} < r_{p23} < r_{23}^\mathrm{max}$, and $r_{31}^\mathrm{min} < r_{p31} < r_{31}^\mathrm{max}$. They are usually computed by explicitly counting triplets of a random distribution of points, but for a box with periodic boundaries these triplet-counts are easy to compute analytically. We follow the approach similar to \citet{2019MNRAS.486L.105P} when computing these triplet counts.

We start by computing a simpler quantity, $RRR^\star\left(r_{12},r_{23}^\mathrm{min},r_{23}^\mathrm{max},r_{31}^\mathrm{min},r_{31}^\mathrm{max}\right)$, the average number of third neighbours for a fixed pair separated by an exact perpendicular distance of $r_{12}$. Fig.~\ref{fig:RRRanalytic} shows the geometry of the problem. For a fixed pair of points, $RRR^\star$ is an average number of points falling within the shaded areas.
\begin{align}
    RRR^\star&\left(r_{12},r_{23}^\mathrm{min},r_{23}^\mathrm{max},r_{31}^\mathrm{min},r_{31}^\mathrm{max}\right) =\\ \nonumber &\overline{\sigma}S^\star(r_{12},r_{23}^\mathrm{min},r_{23}^\mathrm{max},r_{31}^\mathrm{min},r_{31}^\mathrm{max}),
\end{align}
where $\overline{\sigma} = N/L^2$, is the projected density of the points ($N$ being the number of points, and $L$ the side of the cube).

The relationship of this simplified quantity with the full triplet count is,
\begin{align}
    RRR&\left(r_{12}^\mathrm{min},r_{12}^\mathrm{max},r_{23}^\mathrm{min},r_{23}^\mathrm{max},r_{31}^\mathrm{min},r_{31}^\mathrm{max}\right) = \\ \nonumber &\displaystyle\int\limits_{r_{12}^\mathrm{min}}^{r_{12}^\mathrm{max}}\!\!\! RRR^\star\left(r_{12},r_{23}^\mathrm{min},r_{23}^\mathrm{max},r_{31}^\mathrm{min},r_{31}^\mathrm{max}\right) N (\rho 2\pi r_{12}\mathrm{d}r_{12}),
\end{align}
where $N$ is the total number of possible first particles in the triplet and $(\rho 2\pi r_{12}\mathrm{d}r_{12})$ is the average number of second particles in the triplet as we integrate over the $r_{12}$ bin.

 We compute $S^\star$ using the expression for the area of the intersection of two circles
\begin{equation}
\label{circleoverlap}
    \begin{aligned}
        A(d,R,r)&=r^2\arccos{\left(\frac{d^2+r^2-R^2}{2dr}\right)}+R^2\arccos{\left(\frac{d^2-r^2+R^2}{2dR}\right)}\\
         &-\frac{1}{2}\sqrt{(-d+r+R)(d+r-R)(d-r+R)(d+r+R)}.
    \end{aligned}
\end{equation}
Here, $d$ is the distance between the two circles, $R$ and $r$ are the two radii, and $A$ is the shaded area on Fig.~\ref{fig:circleintersect}. From Fig.~\ref{fig:RRRanalytic} and \ref{fig:circleintersect} it is clear that
\begin{align}
    S^\star &= A(r_{12}, r_{23}^\mathrm{max}, r_{31}^\mathrm{max}) - A(r_{12}, r_{23}^\mathrm{max}, r_{31}^\mathrm{min}) \\ \nonumber
    &- A(r_{12}, r_{23}^\mathrm{min}, r_{31}^\mathrm{max}) + A(r_{12}, r_{23}^\mathrm{min}, r_{31}^\mathrm{min}).
\end{align}

In our code we keep track of identical triplets by imposing the condition $r_{p12} < r_{p23} < r_{p31}$. This ensures that we don't count the same physical triplet corresponding to the same particles several times by relabeling the particles 1, 2, and 3. This does not happen (because of the way our code is written) for the triplets for which either three sides or at least two sides of the triangle fall into the same bin, so those triplets are counted more than ones. To correct for this, we apply a permutation factor $N_\mathrm{perm}$ to our $RRR$ counts. The permutation factor is
\begin{equation}
N_{\mathrm{perm}}(r_1,r_2,r_3) = 
\left\lbrace
\begin{array}{lcl}
6   & \mathrm{for}  & r_1\neq r_2\neq r_3,\\
3   & \mathrm{for}  & r_1=r_2\neq r_3\; \mathrm{or}\\
    &               & r_1=r_3\neq r_2 \;\mathrm{or}\\
    &               & r_2=r_3\neq r_1, \\
1   & \mathrm{for}  & r_1=r_2=r_3.
\end{array}\right.
\label{eq:ntilde}
\end{equation}

%%%%%%%%%%%%%%%%%%%%%%%%%%%%%%%%%%%%%%%%%%%%%%%%%%

% Don't change these lines
\bsp	% typesetting comment
\label{lastpage}
\end{document}